\documentclass[aps,prb,reprint,showpacs,superscriptaddress,longbibliography]{revtex4-1}
\usepackage[dvipdfmx]{graphicx}

\usepackage{amsmath,amssymb,amsfonts}
\usepackage{comment}
\usepackage{color}
\usepackage[T1]{fontenc}
\usepackage{textcomp}
\usepackage{hyperref}
\hypersetup{
setpagesize=false,
colorlinks=true,
linkcolor=blue,
citecolor=red,
}
\usepackage{bm,ulem}
\usepackage{mathrsfs}
\usepackage{physics}

\makeatletter
\let\MYcaption\@makecaption
\makeatother
\usepackage{subcaption}
\makeatletter
\let\@makecaption\MYcaption
\makeatother
\begin{document}
\title{Tuning monolayer superconductivity in twisted NbSe$_2$ graphene heterostructures}
\author{Shun Asano}
\email{asano.shun.57x@st.kyoto-u.ac.jp}
\author{Youichi Yanase}
\affiliation{Department of Physics, Kyoto University, Kyoto 606-8502, Japan}
\date{\today}
\begin{abstract}
The recent advent of artificial structures has triggered the emergence of fascinating phenomena that could not exist in natural compounds. A prime example is twisted multilayers, i.e., $\mathrm{moir\acute{e}}$ superlattices represented by magic-angle twisted bilayer graphene (MATBG). As in the case of MATBG, unconventional band hybridization can induce a new type of superconductivity: artificial band engineering by twist induces properties different from the original systems. Here, we apply this perspective to a monolayer superconductor NbSe$_2$ stacked with a twist on doped graphene. We show that the superconducting states of the NbSe$_2$ layer change dramatically by varying the twist angle. Our result shows that twist tuning, in addition to substrate effects, will provide a strategy for designing monolayer superconductors with high controllability.
\end{abstract}
\maketitle
\section{INTRODUCTION}
Exploration of unconventional superconductivity has been a central topic of condensed matter physics for a long time. Lower dimensionality has been thought of as one way to tackle this problem~\cite{Saito2016,Brun2017IOP,Qiu2021AdvMater}. This kind of belief contributed to the development of nanotechnology, and as technology progressed in a top-down direction, many nontrivial phenomena were found by thinning, depositing and interfacing materials~\cite{Ohtomo2004,Gupta2015ProginMater,Gariglio2015PhysicaC,Wang2016IOP,Tan2013NatMater,Liu2012NatCommun,He2013NatMater,Xiang2012PRB}. For example, many researchers have focused on transition metal dichalcogenides (TMD) belonging to van der Waals (vdW) materials~\cite{Mak2010,Ye2012,Voiry2015ChemSocRev,Wang2012NatNanotech,Han2018ChemRev,Devarakonda2020Science,Ding2011Physica}. The vdW materials have layered structures and can be exfoliated easily. This allowed for controlling on the atomic scale, which led to investigation and clarification of monolayer superconductivity.

Some hexagonal TMD such as NbSe$_2$, TaS$_2$, and doped MoS$_2$ are known to exhibit superconductivity from bulk to even monolayer~\cite{Ye2012,Xi2015NatNanotech,Costanzo2016NatNanotech,Saito2016NatPhys,Moratalla2016NatCommun,Peng2018NanoLett}. In particular, monolayer NbSe$_2$ is intrinsically metallic and becomes a multigap superconductor with Fermi surfaces around the K valley and around the $\Gamma$ point~\cite{Yokoya2001Science,Khestanova2018NanoLett,Dvir2018NatCommun,Wickramaratne2020PRX,Noat2015PRB,Zheng2019PRB}. In addition, hexagonal TMD have in common a strong spin-orbit coupling (SOC), so-called Ising SOC, in the monolayer limit due to the spatial inversion symmetry breaking. This causes spin-momentum locking, and therefore the superconducting (SC) pairing is extremely robust to external in-plane magnetic fields~\cite{Xi2016NatPhys,Saito2016NatPhys,Lu2015Science,Wang2017NatCommun,Ugeda2016NatPhys,Barrera2018NatCommun,Xing2017NanoLett,Wan2022AdvMater,Moratalla2016NatCommun,Costanzo2016NatNanotech}. The robustness of such a SC state to magnetic fields makes them crucial both from a fundamental point of view and for possible applications.

On the other hand, subsequently, many attempts have been made in a bottom-up direction recently, i.e., stacking materials to create a superlattice by replacing, alternating, or twisting layers~\cite{Naritsuka2021IOP,Goh2012PRL,Shishido2010Science,Mizukami2011NatPhys,Niu2015ProgSurfSci,Kim2016NanoLett,Veneri2022PRB,Nie2023NanoLett,Li2019PRB,Xiao2023PRB,Dreher2021NanoLett,Geim2013Nature,Kennes2021NatPhys,Behura2021EmergeMater,Liao2019AppMater,Tran20212DMater}. Notably, there has been intense research on $\mathrm{moir\acute{e}}$ physics in twisted vdW bilayers in the last five years. In 2018 discovery of superconductivity in magic-angle twisted bilayer graphene (MATBG) marked a milestone~\cite{Cao2018Nature1,Cao2018Nature2}, which drew a great deal of attention to the superconductivity as the first emerging by the twist~\cite{Matthew2019Science,Tarnopolsky2019PRL,Andrei2020NatMater,Ramires2018PRL,Qin2021PRL,Ray2019PRB,Scheurer2020PRR,Peltonen2018PRB,Christos2023NatCommun,Oh2021Nature,Wu2018PRL,Po2018PRX,Isobe2018PRX,Saito2020NatPhys,Arora2020Nature,Gonzalez2019PRL,Emilio2019SciAdv,Kerelsky2019Nature}. As represented by the flat band in MATBG, the most valuable aspect of the $\mathrm{moir\acute{e}}$ structures is the control of band hybridization by the twist angle~\cite{Lisi2021NatPhys,Morell2010PRB,Zhang2020NatPhys,Devakul2021NatCommun,Xian2021NanoLett}. This means that artificial band engineering is possible with the control of experimental parameters. Furthermore, twisted heterostructures are expected to induce novel physical properties due to a wider range of material selection compared to $\mathrm{moir\acute{e}}$ structures of the same materials~\cite{Veneri2022PRB,Nie2023NanoLett,Li2019PRB,Xiao2023PRB,Dreher2021NanoLett,Geim2013Nature,Kennes2021NatPhys,Behura2021EmergeMater,Liao2019AppMater,Tran20212DMater}. This indicates that electronic states suitable for unusual superconductivity could be extracted by combining selected properties of different materials.

As previous studies without a twist, a monolayer FeSe on SrTiO$_{3}$ substrate has been found to exemplify high-temperature superconductivity by the combination of properties of FeSe monolayer and the substrate effect~\cite{Tan2013NatMater,Liu2012NatCommun,He2013NatMater,Xiang2012PRB}. Nevertheless, the modulation of the intrinsic SC property is not yet well understood in twisted heterostructures of monolayer superconductors and other materials. A monolayer superconductor stacked with a twist on a monolayer substrate is regarded as a twisted bilayer, in which the monolayer SC state is expected to be strongly tuned by the substrate. Especially when the substrate is metallic, band hybridization near the Fermi level could modulate the monolayer SC states. In addition, the presence of an experimental parameter, i.e., the twist angle, can make it possible to design physical properties infeasible so far. Therefore, tuning monolayer SC states enabled by substrate effects and twist has the potential to induce unconventional phenomena on a monolayer superconductor such as the enhancement of transition temperature, topological superconductivity, and SC diode effect~\cite{Tan2013NatMater,Liu2012NatCommun,He2013NatMater,Xiang2012PRB,Zhang2018Science,Machida2019,He2018CommunPhys,Kezilebieke2020Nature,Xie2023PRL}.

In this paper, we investigate the property of monolayer superconductivity, especially in a monolayer NbSe$_2$ stacked with a twist on doped graphene. The doped graphene is treated as a metallic substrate. We then regard this system as a twisted bilayer and identify each as a top and bottom layer. The previous work revealed that there are two scenarios of band hybridization because of a large mismatch of lattice constants and significant difference in the size of Fermi surfaces in twisted NbSe$_2$ graphene heterostructures~\cite{Gani2019PRB}. Moreover, they showed that SC pairing could be induced into the graphene layer by proximity effect, but the details of SC states in the NbSe$_2$ layer are not paid attention to and uncovered. In this paper, we focus on the property of the NbSe$_2$ layer and reveal its twist angle dependence. In $\mathrm{moir\acute{e}}$ superlattices, interlayer band hybridization and long periodicity in real space can affect superconductivity. We mainly focus on the former effect and construct an effective model based on which we theoretically treat the multigap superconductivity for arbitrary twist angles.

\section{FORMULATION}
First, we explain the formulation of this work.
The previous work~\cite{Gani2019PRB} mainly focused on graphene and adopted a continuum model~\cite{Mele2010PRB,Rafi2011PNAS,Koshino2020PRB,Santos2007PRL}. However, due to the large lattice mismatch, it is not adequate for the analysis of the angle dependence of SC states. Alternatively, we here consider a generalized Umklapp process for general commensurate or incommensurate bilayers~\cite{Koshino2015IOP,David2019PRB}. Then, we build a model focusing on the low-energy states because only the low-energy electronic states near the Fermi level are important for superconductivity.

\subsection{Low-energy effective model of each layer}
Before constructing an entire Hamiltonian describing NbSe$_2$ graphene heterostructures, we introduce the electronic structure of each layer and show the intralayer Hamiltonian for the low-energy effective model. In this paper, we use the parameters evaluated in Ref.~\onlinecite{Gani2019PRB}.

We write the primitive lattice vectors as $\mathbf{a}_i\,(i=1,2)$ for the NbSe$_2$ layer and $\mathbf{\tilde{a}}_i$ for the graphene layer, which are all oriented in-plane direction. The reciprocal lattice vectors are written as $\mathbf{G}_j$ and $\mathbf{\tilde{G}}_j$ for NbSe$_2$ and graphene layers respectively, to satisfy $\mathbf{a}_i\cdot\mathbf{G}_j=\mathbf{\tilde{a}}_i\cdot\mathbf{\tilde{G}}_j=2\pi\delta_{ij}$. In the same way, other physical quantities of NbSe$_2$ and graphene layers shall be distinguished with or without a tilde.

In graphene, the carbon atoms form a honeycomb lattice with two sublattices, which are specified by $X=A, B$. The lattice constant is $\tilde{a}$=2.46 \AA. Here, we choose the primitive lattice vectors $\mathbf{\tilde{a}}_1=\tilde{a}[-1/2, \sqrt{3}/2]$, $\mathbf{\tilde{a}}_2=\tilde{a}[-1/2, -\sqrt{3}/2]$ and primitive reciprocal lattice vectors $\mathbf{\tilde{G}}_1=\frac{4\pi}{\sqrt{3}\tilde{a}}[-\sqrt{3}/2, 1/2]$ and $\mathbf{\tilde{G}}_2=\frac{4\pi}{\sqrt{3}\tilde{a}}[-\sqrt{3}/2, -1/2]$. Thus, the lattice positions are given by
\begin{equation}
\mathbf{\tilde{R}}_{\tilde{X}}=\tilde{n}_1\mathbf{\tilde{a}}_1+\tilde{n}_2\mathbf{\tilde{a}}_2+\mathbf{\tilde{d}}_{X},
\label{eq1}
\end{equation}
where $\tilde{n}_i$ are integers and $\mathbf{\tilde{d}}_{A}=[0,0]$, $\mathbf{\tilde{d}}_{B}=[0, -\tilde{a}/\sqrt{3}]$ are positions of $A,B$ sublattices. 

The low-energy states are located near the valley points of Brillouin zone (BZ) $\mathbf{\tilde{K}}_{\tilde{\xi}}=[\tilde{\xi}\frac{4\pi}{3\tilde{a}}, 0]$, where $\tilde{\xi}=\pm1$ labels the valley degree of freedom (DOF). The single-particle Hamiltonian around these valleys is described by massless Dirac fermions
\begin{align}
    \mathcal{H}^{(\tilde{\xi})}_{G}=&\sum_{\mathbf{\tilde{p}}, X, X',\tilde{\sigma},\tilde{\sigma}'} c^{\dagger}_{\mathbf{\tilde{K}}_{\tilde{\xi}}+\mathbf{\tilde{p}}, X', \tilde{\sigma}'}
    H_{G}^{(\tilde{\xi})} (\mathbf{\tilde{p}})
    \,c_{\mathbf{\tilde{K}}_{\tilde{\xi}}+\mathbf{\tilde{p}}, X, \tilde{\sigma}}, 
    \label{eq2}
    \\
    H_{G}^{(\tilde{\xi})} (\mathbf{\tilde{p}})=&\left[\hbar v_{\rm F}(\tilde{\xi}\tau_x \tilde{p}_x+\tau_y \tilde{p}_y)-\mu_g\right] \tilde{\sigma}_0, 
    \label{eq3}
\end{align}
where $\tilde{\sigma},\tilde{\sigma}'=\pm1$ are the labels of up spin and down spin, $\mathbf{\tilde{p}}$ is the wave vector measured from each valley point, $\tilde{\sigma}_0=\delta_{\tilde{\sigma},\tilde{\sigma}'}$ means an identity matrix in spin space, $v_{\rm F}$ is Fermi velocity, $\mu_g$ is a chemical potential of graphene, and $\tau_x, \tau_y$ are Pauli matrices representing the sublattice DOF. $c^{\dagger}_{\mathbf{\tilde{k}}, X, \tilde{\sigma}}$ ($c_{\mathbf{\tilde{k}}, X, \tilde{\sigma}}$) is the creation (annihilation) operator of a low-energy electron around each valley. It is convenient to define the following spinor for each valley:
\begin{equation}
    \mathbf{C}^{\dagger}_{ \mathbf{\tilde{p}}, \tilde{\xi}}= \left[c^{\dagger}_{\mathbf{\tilde{K}}_{\tilde{\xi}}+\mathbf{\tilde{p}},A,\uparrow}, c^{\dagger}_{\mathbf{\tilde{K}}_{\tilde{\xi}}+\mathbf{\tilde{p}},A,\downarrow}, c^{\dagger}_{\mathbf{\tilde{K}}_{\tilde{\xi}}+\mathbf{\tilde{p}},B,\uparrow}, c^{\dagger}_{\mathbf{\tilde{K}}_{\tilde{\xi}}+\mathbf{\tilde{p}},B,\downarrow} \right].
    \label{eq4}
\end{equation}

Similarly, the H-type NbSe$_2$ also forms a honeycomb lattice with which a sublattice consists of one Nb atom and two Se atoms. The Se atoms symmetrically displace above and below the plane formed by the Nb atoms (Fig.~\ref{fig1}). The lattice constant is $a$=3.48 \AA\,  and primitive lattice vectors $\mathbf{a}_i$, primitive reciprocal lattice vectors $\mathbf{G}_i$, and positions of the K valley points $\mathbf{K}_{\xi}$ are obtained by replacing $\tilde{a} \rightarrow a$ in those of graphene. The BZ is also hexagonal, and in contrast to graphene the Ising SOC results in a spin splitting of Fermi Surfaces (FSs), which conserve spin $\sigma_z$. 

Unlike graphene, the low-energy states are located around both the K valleys and the $\Gamma$ point. In the monolayer limit, only the orbitals of Nb atoms contribute to the low-energy states. This fact allows us to consider only the Nb sublattice DOF as long as our focus is limited to the low-energy region. The effective models for electrons around the $\Gamma$ point and the $\mathbf{K}_{\xi} \, (\xi=\pm1)$ valleys are as follows:
\begin{align}
    \mathcal{H}_{\Gamma}=&\sum_{\mathbf{p}, \sigma, \sigma'} d^{\dagger}_{\mathbf{p}, \sigma'} H_{\Gamma}(\mathbf{p}) \, d_{\mathbf{p}, \sigma}, 
    \label{eq5}
    \\ \mathcal{H}_{\mathbf{K}}^{(\xi)}=&\sum_{\mathbf{p},\sigma,\sigma'}d^{\dagger}_{\mathbf{K}_{\xi}+\mathbf{p}, \sigma'} H_{\mathbf{K}}^{(\xi)}(\mathbf{p}) \, d_{\mathbf{K}_{\xi}+\mathbf{p}, \sigma},
    \label{eq6}
    \\
    H_{\Gamma}(\mathbf{p})=&\,\epsilon_1(\mathbf{p}; \eta_{\Gamma}, \lambda_{\Gamma})\sigma_0+\epsilon_2(\mathbf{p}; \nu_{\Gamma})\sigma_z,
    \label{eq7}
    \\
    H_{\mathbf{K}}^{(\xi)}(\mathbf{p})=&\,\epsilon_1(\mathbf{p}; \eta_{K}, \lambda_{K})\sigma_0+\xi\epsilon_2(\mathbf{p};\nu_{K})\sigma_0 {\notag}
    \\
    &+\xi\epsilon_1(\mathbf{p};\eta'_{K},\lambda'_{K})\sigma_z, 
    \label{eq8}
\end{align}
where 
\begin{align}
    \epsilon_1(\mathbf{p};\eta,\lambda)=&\eta+\lambda(p^2_x+p^2_y),
    \label{eq9}
    \\
    \epsilon_2(\mathbf{p}; \nu)=&\nu (2p^2_x-6p_x p^2_y).
    \label{eq10}
\end{align}

Like the Hamiltonian of graphene, $\mathbf{p}$ is measured from the $\Gamma$ or $\mathbf{K}_{\xi}$ points.
Here, $d^{\dagger}_{\mathbf{k},\sigma}$ ($d_{\mathbf{k},\sigma}$) is the creation (annihilation) operator for an electron on NbSe$_2$. It is worth mentioning again that we ignore the sublattice DOF of Se atoms. We also define the following spinors for NbSe$_2$ low-energy states:
\begin{align}
    \mathbf{D}^{\dagger}_{\mathbf{p}}=&\left[d^{\dagger}_{\mathbf{p},\uparrow}, d^{\dagger}_{\mathbf{p},\downarrow}\right],
    \label{eq11}
    \\
    \mathbf{D}^{\dagger}_{\mathbf{p},\xi}=&\left[d^{\dagger}_{\mathbf{K}_{\xi}+\mathbf{p},\uparrow},d^{\dagger}_{\mathbf{K}_{\xi}+\mathbf{p},\downarrow}\right].
    \label{eq12}
\end{align}

Hereafter, all wave vectors are normalized by the unit length in reciprocal space of NbSe$_2$ $2\pi/a$.  
The parameters of the above effective models are shown in Table~\ref{Table1}. 
Note that graphene is doped in the heterostructure ($\mu_g \ne 0$)~\cite{Gani2019PRB}, making the graphene metallic substrate. 

\begin{table*}[ht]
\caption{Parameters of the intralayer Hamiltonian~\cite{Gani2019PRB}. The energy unit is eV.}
\centering
\renewcommand\arraystretch{2}
\begin{tabular*}{18cm}{@{\extracolsep{\fill}}ccccccccc}
\hline \hline
 $\mu_g$ & $\eta_{\Gamma}$ & $\lambda_{\Gamma}$ & $\nu_{\Gamma}$ & $\eta_{K}$ & $\lambda_{K}$ & $\nu_{K}$ & $\eta'_{K}$ & $\lambda'_{K}$\\
\hline
$-0.4$&0.5641&$-7.0640$ $[a/(2\pi)]^2$&0.5085 $[a/(2\pi)]^3$&0.4526&$-9.0940$ $[a/(2\pi)]^2$&3.07 $[a/(2\pi)]^3$&0.0707&-0.33 $[a/(2\pi)]^2$\\
\hline \hline
\end{tabular*}
\label{Table1}
\end{table*}
 
\subsection{General theory for misoriented bilayers}
\begin{figure}[b]
    \hspace*{-0.0cm}
    \includegraphics[width=8.5cm]{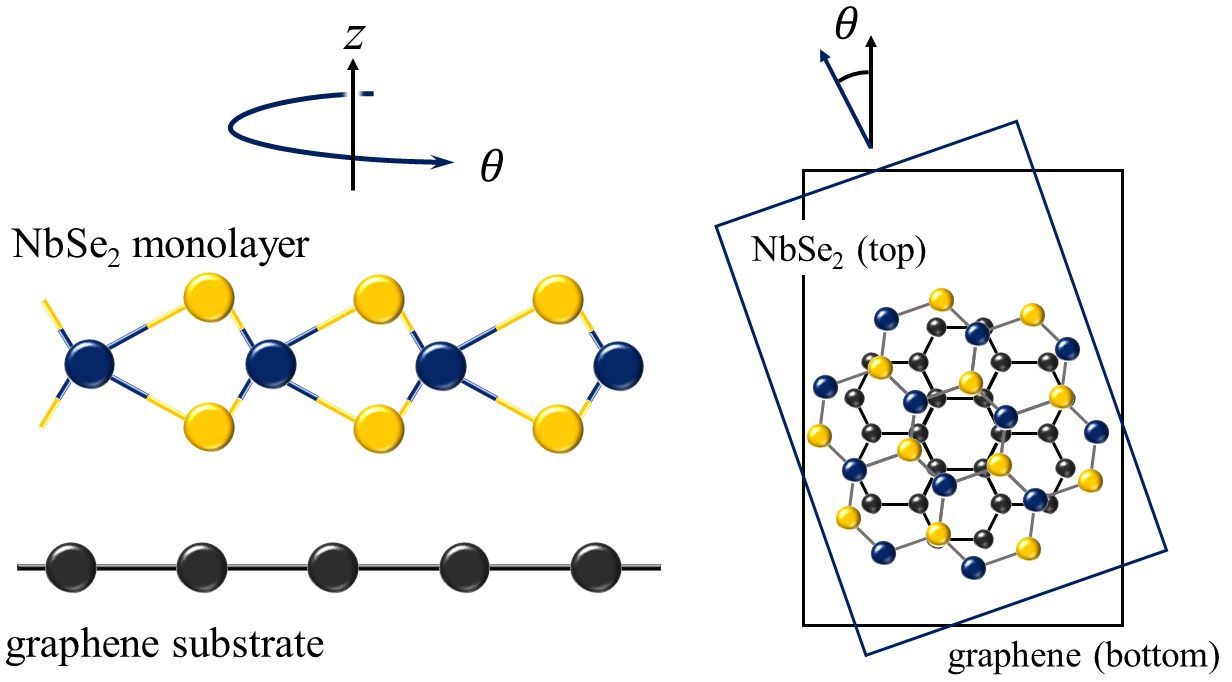}
    \caption{A schematic figure for the twisted NbSe$_2$ graphene bilayer. The left (right) panel shows the side (top) view. The black, blue, and yellow spheres represent carbon, Nb, and Se atoms, respectively. The top NbSe$_2$ layer is twisted counterclockwise with respect to the bottom graphene layer.}
    \label{fig1}
\end{figure}

Figure~\ref{fig1}  shows the considered setup in this paper. The monolayer NbSe$_2$ is stacked on graphene, and the top layer is twisted counterclockwise by $\theta$ relative to the bottom layer. This means at the same time that the graphene layer is twisted by $-\theta$ to the NbSe$_2$ layer. Hence, we regard this system as one in which the NbSe$_2$ layer is fixed and the graphene layer is twisted in the clockwise direction with twist angle $\theta$. Accordingly, vectors of the graphene layer in both real and reciprocal space are transformed such as $\mathbf{\tilde{a}}_i \rightarrow R(-\theta) \mathbf{\tilde{a}}_i$, $\mathbf{\tilde{G}}_i \rightarrow R(-\theta) \mathbf{\tilde{G}}_i$, where
 \begin{equation}
     R(\theta)=
     \begin{bmatrix}
         \cos{\theta} & -\sin{\theta} \\
         \sin{\theta} & \cos{\theta}
     \end{bmatrix},
     \label{eq13}
 \end{equation}
is the rotation matrix and we use $\left(R^{\dagger}\right)^{-1}=R$. Other vectors follow the same transformation, and in the following, characters with tildes implicitly represent vectors undergone a rotational transformation.
 
The entire Hamiltonian for twisted heterostructures is the sum of the above intralayer Hamiltonian and an interlayer hopping term, written as $\mathcal{H}=\mathcal{H}_{\mathrm{G}}+\mathcal{H}_{\mathrm{NbSe_2}}+\mathcal{H}_T$. $\mathcal{H}_{\mathrm{G}}$ and $ \mathcal{H}_{\mathrm{NbSe_2}}$ are the intralayer Hamiltonian of graphene and NbSe$_2$ mentioned in the previous subsection.

The interlayer coupling is small enough to be treated as a perturbation compared to the energy scales of $\mathcal{H}_{\mathrm{G}}, \mathcal{H}_{\mathrm{NbSe_2}}$ in this system. This is a consequence of the large lattice mismatch, in addition to the fact that bilayers are connected by weak vdW forces~\cite{Gani2019PRB,Santos2007PRL,Zhang2014PRL}. Therefore, we adopt a theory for general misoriented bilayers~\cite{Koshino2015IOP}. In this scheme, an electronic state with a Bloch wave vector $\mathbf{k}$ of the NbSe$_2$ layer and that of the graphene layer with $\mathbf{\tilde{k}}$ are coupled only when $\mathbf{k+G}=\mathbf{\Tilde{k}}+\mathbf{\Tilde{G}}$ where $\mathbf{G}=m_1\mathbf{G}_1+m_2\mathbf{G}_2$ and $\mathbf{\tilde{G}}=\tilde{m_1}\mathbf{\tilde{G}}_1+\tilde{m_2}\mathbf{\tilde{G}}_2$ are reciprocal lattice vectors of each layer. In other words, the neighboring layers' states are coupled at common wave vectors in the extended BZ scheme. Instead, however, it could be considered that the states of graphene with $\mathbf{\tilde{k}}=\mathbf{k}+\mathbf{G}-\mathbf{\tilde{G}}$ couple to the states with $\mathbf{k}$ in the first BZ of NbSe$_2$ in the reduced zone scheme. Thus, we construct a model by adopting the latter interpretation.
 
The interlayer Hamiltonian is written as
 \begin{equation}
\mathcal{H}_T=\sum_{\mathbf{k},\mathbf{\tilde{k}}}d^{\dagger}_{\mathbf{k},\sigma} T_X(\mathbf{k},\mathbf{\tilde{k}})c_{\mathbf{\tilde{k}},X,\tilde{\sigma}}+h.c.,
\label{eq14}
 \end{equation}
where
\begin{equation}
T_X(\mathbf{k},\mathbf{\tilde{k}})=\sum_{\mathbf{G},\mathbf{\tilde{G}}} t(\mathbf{k}+\mathbf{G}) e^{i\mathbf{\tilde{G}}\cdot \mathbf{\tilde{d}}_X} \delta_{\sigma,\tilde{\sigma}}\delta_{\mathbf{k}+\mathbf{G},\mathbf{\tilde{k}}+\mathbf{\tilde{G}}},
\label{eq15}
\end{equation}
and $t(\mathbf{q})$ is the amplitude of the coupling which decays dramatically in large $\mathbf{q}$. Therefore, we merely consider a limited number of components in the summation of Eq.~\eqref{eq15} and ignore graphene valleys except for the innermost valleys in the extended zone scheme.
 
\begin{figure*}[t]
    \begin{minipage}{0.45\linewidth}
    \hspace*{-0.6cm}
    \includegraphics[width=9cm]{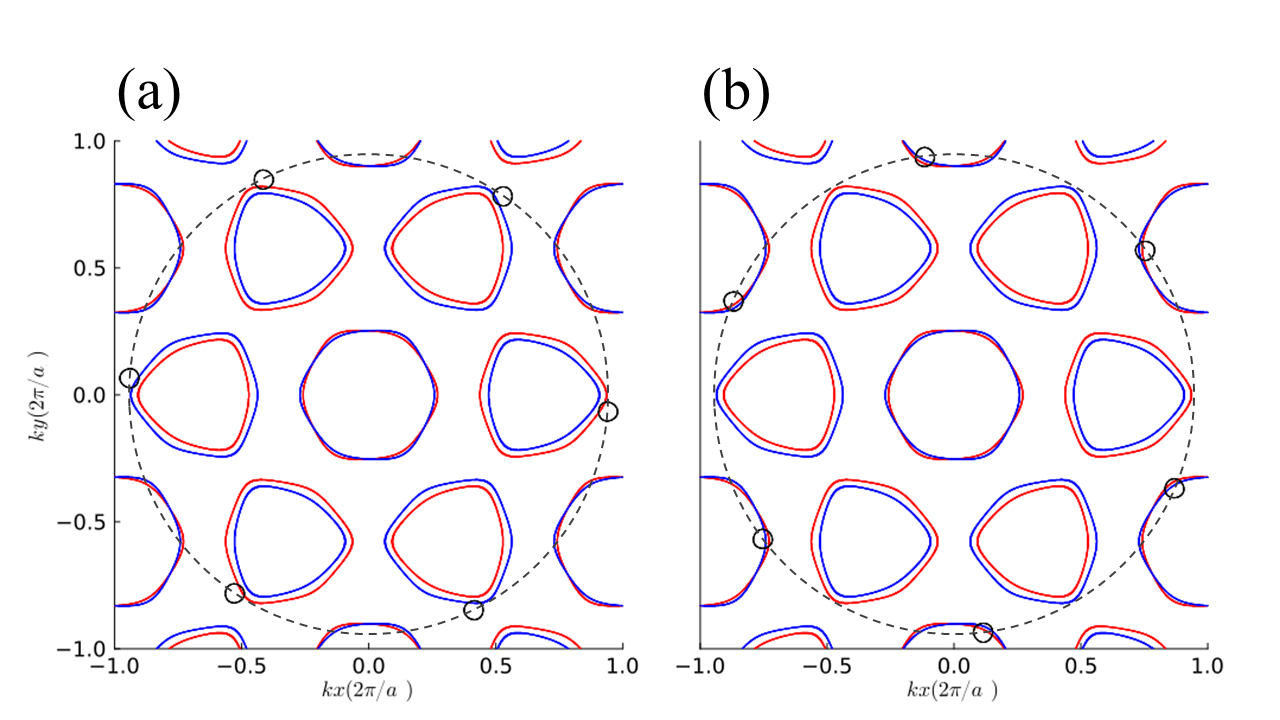}
    \end{minipage}
    \begin{minipage}{0.45\linewidth}
    \centering
    \includegraphics[width=9cm]{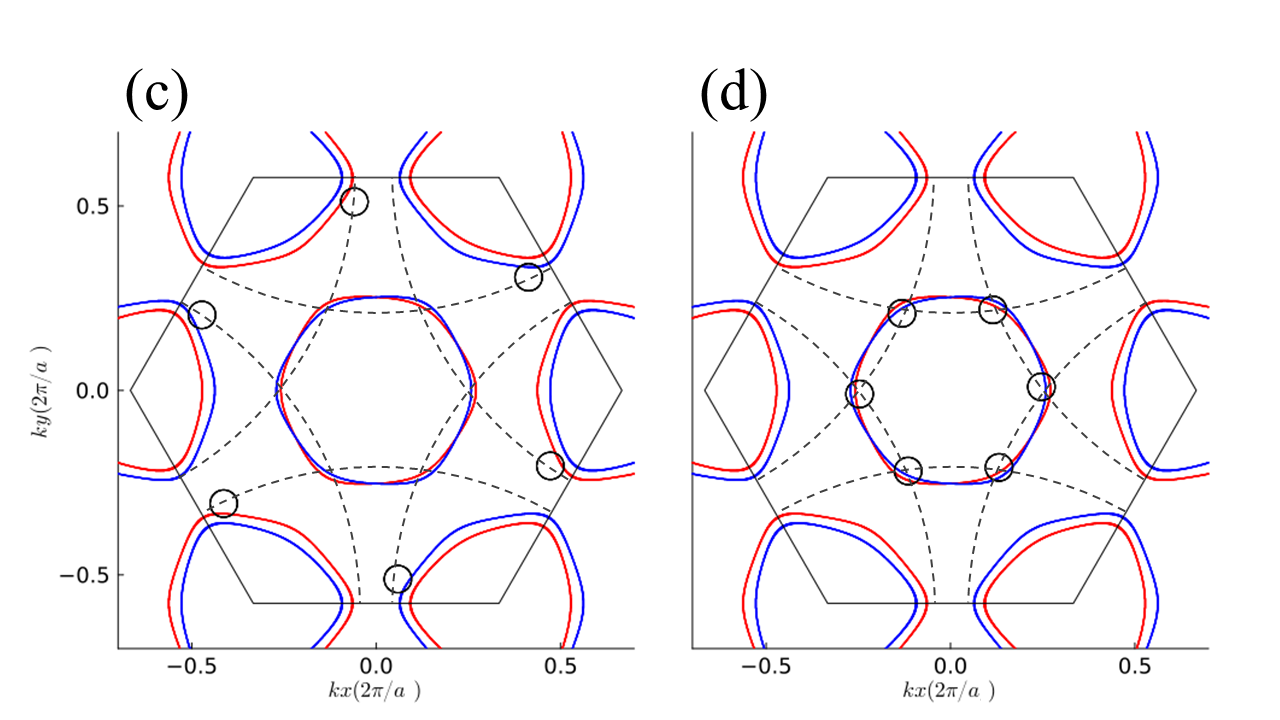}
    \end{minipage}
    \caption{Relative position of FSs on graphene and NbSe$_2$ layers in the extended zone scheme (a), (b), and the reduced zone scheme (c), (d). The red (blue) lines represent the FSs of NbSe$_2$ with spin up (down), and the black circles show the graphene FSs. We also depict dashed lines which are traces of graphene's Dirac points. In (a) and (c), the twist angle is $\theta=4^{\circ}$ where the FSs around the K valleys hybridize. By contrast, in (b) and (d), the twist angle is $\theta=23^{\circ}$, where the graphene FSs hybridize the NbSe$_2$ FSs around the $\Gamma$ point. Note that we use a tight-binding model for depicting the FSs of NbSe$_2$.~\cite{Sticlet2019PRB}}
    \label{fig2}   
\end{figure*}

Figure~\ref{fig2} illustrates the FSs of layers in the extended zone scheme and corresponding positions inside the first BZ of NbSe$_2$ for $\theta=4^{\circ},23^{\circ}$, for example. There are two scenarios of hybridization between the graphene and NbSe$_2$ FSs. When the twist angle is small $\theta \simeq 0^{\circ}$, the graphene and NbSe$_2$ FSs around each K valley hybridize.  On the other hand, the graphene FSs around valleys hybridize the NbSe$_2$ FS around the $\Gamma$ point, when the twist angle is large $\theta \simeq 30^{\circ}$. Owing to the symmetry of reciprocal space, we can restrict the twist angle as $0^{\circ} \leq \theta \leq 60^{\circ}$.  

First, we consider the case of a small twist angle $\theta$ around $0^{\circ}$. In this case, valleys of the same sign in each layer are hybridized, i.e., $\xi=\tilde{\xi}$. 
Let us ignore the FS around the $\Gamma$ point of NbSe$_2$ for simplicity because it is hardly affected by band hybridization.  Under this assumption, we can describe the low-energy states around the valley of each sign simply by considering the Bloch basis of NbSe$_2$ and three equivalent bases for graphene. In addition, the offset vectors between $\tilde{\mathbf{K}}_{\xi}$ and $\mathbf{K}_{\xi}$ are naturally derived from Eq.~\eqref{eq15}. The summation over ${\tilde{\mathbf{G}}}$ is restricted to three vectors $\mathbf{\tilde{g}}_{1,\xi}=\mathbf{0}$, $\mathbf{\tilde{g}}_{2,\xi}=\xi \tilde{\mathbf{G}}_1$, and $\mathbf{\tilde{g}}_{3,\xi}=\xi \tilde{\mathbf{G}}_2$. Similarly, the sum over $\mathbf{G}$ could be restricted as $\mathbf{g}_{1,\xi}=\mathbf{0}$, $\mathbf{{g}}_{2,\xi}=\xi \mathbf{G}_1$, and $\mathbf{g}_{3,\xi}=\xi \mathbf{G}_2$. Therefore, the three offset vectors $\mathbf{q}_{j,\xi}$ $(j=1,2,3)$ for each valley become as follows:
 \begin{equation}
     \mathbf{q}_{j,\xi}=\mathbf{K}_{\xi}-\mathbf{\tilde{K}}_{\xi}+\mathbf{g}_{j,\xi}-\mathbf{\tilde{g}}_{j,\xi}.
     \label{eq16}
 \end{equation}
Because of the relationship $\mathbf{q}_{j,-}=-\mathbf{q}_{j,+}$, we can rewrite $\mathbf{q}_{j,\xi}=\xi \mathbf{q}_j$ by defining $\mathbf{q}_j=\mathbf{q}_{j,+}$. It should be noticed that $\mathbf{q}_j$ depend on the twist angle $\theta$.
\begin{figure}[t]
    \centering
    \includegraphics[height=6cm]{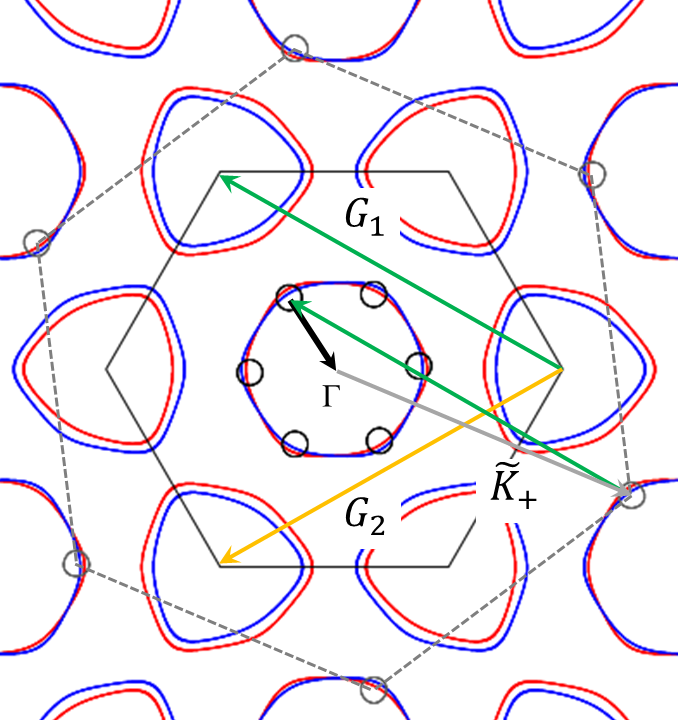}
    \caption{Illustration of a offset vector for $\theta=23^{\circ}$. The green and yellow arrows are the reciprocal lattice vectors of NbSe$_2$, and the gray arrow shows the valley point of graphene $\tilde{K}_{+}$. The solid and dashed hexagons represent the BZs of NbSe$_2$ and graphene, respectively. We depict one of the offset vectors $\mathbf{q}_{1}$ in Eq.~\eqref{eq22} with a black arrow.}
    \label{fig3}
\end{figure}

The wave number $\mathbf{k}$ in NbSe$_2$ is coupled to $\tilde{\mathbf{k}}=\mathbf{k}+\mathbf{G}-\tilde{\mathbf{G}}$ in graphene as mentioned above. Hence, $\mathbf{p}$ in NbSe$_2$ and $\mathbf{\tilde{p}}=\mathbf{p}+\mathbf{q}_{j,\xi}$ in graphene are coupled, where $\mathbf{p}$ is the wave vector measured from $\mathbf{K}_{+}$ or $\mathbf{K}_{-}$. The interlayer Hamiltonian between these states is written as~\cite{Zhang2014PRL}
 \begin{equation}
     \mathcal{H}_T=\sum_{\mathbf{p},\xi} \sum_{j=1}^{3} \mathbf{D}^{\dagger}_{\mathbf{p},\xi} T_{j,\xi} \mathbf{C}_{\mathbf{p}+\xi\mathbf{q}_j,\xi}+h.c.,
     \label{eq17}
 \end{equation}
where 
\begin{equation}
\begin{aligned}
T_{1,\xi}=&
    \begin{bmatrix}
        t & 0 & t & 0 \\
        0 & t & 0 & t
    \end{bmatrix},
\\
T_{2,\xi}=&
    \begin{bmatrix}
        t & 0 & t e^{-i \xi (2\pi)/3} & 0 \\
        0 & t & 0 & t  e^{-i \xi (2\pi)/3}
    \end{bmatrix},
\\
T_{3,\xi}=&
    \begin{bmatrix}
        t & 0 & t e^{i \xi (2\pi)/3} & 0 \\
        0 & t & 0 & t  e^{i \xi (2\pi)/3}
    \end{bmatrix},
\end{aligned}
\label{eq18}
\end{equation}
and we here use $\mathbf{\tilde{g}}_1\cdot \mathbf{\tilde{d}}_B=0$, $\mathbf{\tilde{g}}_2\cdot \mathbf{\tilde{d}}_B=-2\pi/3$, $\mathbf{\tilde{g}}_3\cdot \mathbf{\tilde{d}}_B=2\pi/3$.
In this paper, we adopt $t=20$~meV as an interlayer hopping energy~\cite{Gani2019PRB}. Therefore, the entire effective Hamiltonian of the twisted heterostructure describing the low-energy states around each $\mathbf{K}_{\xi}$ valley is obtained as 
\begin{equation}
    \mathcal{H}_{K-K}^{(\xi)}=\sum_{\mathbf{p}} \psi_{K,\xi}^{\dagger}(\mathbf{p}) H_{K-K}^{(\xi)} (\mathbf{p}) \psi_{K,\xi}(\mathbf{p}),
    \label{eq19}
\end{equation}
where
\begin{widetext}
\begin{equation}
    H_{K-K}^{(\xi)}(\mathbf{p})=\left[
    \begin{array}{ccccccc}
        H_{\mathbf{K}}^{(\xi)}(\mathbf{p}) & T_{1,\xi} & T_{2,\xi} & T_{3,\xi} \\ 
        T_{1,\xi}^{\dagger} & H_{G}^{(\xi)} (\mathbf{p}+\xi\mathbf{q}_1) & \mathbf{0} & \mathbf{0} \\
        T_{2,\xi}^{\dagger} & \mathbf{0} & H_{G}^{(\xi)} (\mathbf{p}+\xi\mathbf{q}_2) & \mathbf{0} \\
        T_{3,\xi}^{\dagger} & \mathbf{0} & \mathbf{0} & H_{G}^{(\xi)} (\mathbf{p}+\xi\mathbf{q}_3)\\
    \end{array}
    \label{eq20}
    \right],
\end{equation}
\end{widetext}
and we define the following spinor for simple notation:
\begin{equation}
    \psi_{K,\xi}^{\dagger}(\mathbf{p})= \left[\mathbf{D}_{\mathbf{p},\xi}^{\dagger}, \mathbf{C}^{\dagger}_{\mathbf{p}+\xi\mathbf{q}_1,\xi}, 
    \mathbf{C}^{\dagger}_{\mathbf{p}+\xi\mathbf{q}_2,\xi},
    \mathbf{C}^{\dagger}_{\mathbf{p}+\xi\mathbf{q}_3,\xi}\right].
    \label{eq21}
\end{equation}

Next, we consider the case where $\theta$ is around $30^{\circ}$. The interlayer Hamiltonian can be formulated in the same way as in the former case, but the two K valleys of graphene can not be considered independently because both valleys hybridize the FS of NbSe$_2$ around the $\Gamma$ point. Moreover, we have to redefine the three offset vectors. In this case, $\tilde{\mathbf{g}}_{j,\tilde{\xi}}$ are the same as those of the aforementioned case. However, the restricted $\mathbf{G}$ change as $\mathbf{g}_{1,\tilde{\xi}}=-\tilde{\xi}\mathbf{G}_1$, $\mathbf{g}_{2,\tilde{\xi}}=\tilde{\xi}(\mathbf{G}_1-\mathbf{G}_2)$, and $\mathbf{g}_{3,\tilde{\xi}}=\tilde{\xi}\mathbf{G}_2$, and the offset vectors between $\tilde{\mathbf{K}}_{\tilde{\xi}}$ and $\Gamma$ become $\mathbf{q}_{j,\tilde{\xi}}=\tilde{\xi}\mathbf{q}_j$ where  
\begin{equation}
    \mathbf{q}_{j}=-\tilde{\mathbf{K}}_{+}+\mathbf{g}_{j,+}-\mathbf{\tilde{g}}_{j,+}.
    \label{eq22}
\end{equation}

\begin{figure}[t]
    \centering
    \includegraphics[width=5cm]{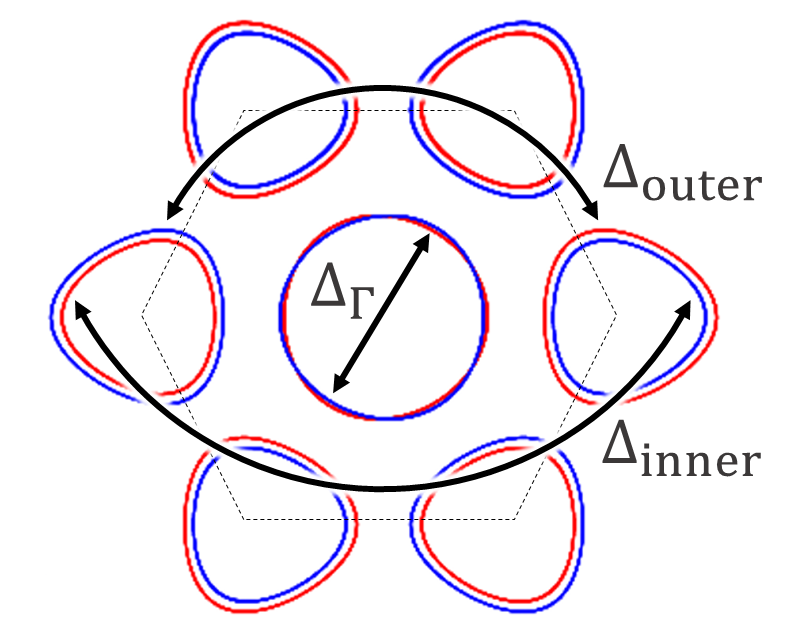}
    \caption{Illustration of three order parameters $\Delta_{\Gamma}$, $\Delta_{\mathrm{inner}}$, and $\Delta_{\mathrm{outer}}$ on the NbSe$_2$ layer. Here we consider simple s-wave Cooper pairing. Due to spin splitting, two order parameters $\Delta_{\mathrm{inner}}$ and $\Delta_{\mathrm{outer}}$ can be considered on the inner and outer FSs around the K valleys, corresponding to the parity mixing in noncentrosymmetric superconductors~\cite{Bauer2012}.}
    \label{fig4}
\end{figure}

For example, Fig.~\ref{fig3} shows the offset vector in addition to other vectors for $\theta=23^{\circ}$. We depict a representative offset vector $\mathbf{q}_{1}$ by the black arrow. In Fig.~\ref{fig3}, we see the necessity for considering the two valley FSs of graphene. Thus, the basis function is spanned by those of NbSe$_2$ and valleys of graphene with $\tilde{\xi} = \pm 1$. The entire Hamiltonian for the low-energy states around the $\Gamma$ point is 
\begin{equation}
    \mathcal{H}_{K-\Gamma}=\sum_{\mathbf{p}}\psi_{\Gamma}^{\dagger}(\mathbf{p}) H_{K-\Gamma}(\mathbf{p})\psi_{\Gamma}(\mathbf{p}),
    \label{eq23}
\end{equation}
where $ H_{K-\Gamma}(\mathbf{p})$ and $\psi_{\Gamma}^{\dagger}(\mathbf{p})$ are the following matrix and spinor:
\begin{widetext}
\begin{equation}
    H_{K-\Gamma}(\mathbf{p})=\left[
    \begin{array}{ccccccc}
        H_{\mathbf{\Gamma}}(\mathbf{p}) & T_{1,+} & T_{2,+} & T_{3,+} & T_{1,-} & T_{2,-} & T_{3,-} \\ 
        T_{1,+}^{\dagger} & H_{G}^{(+)} (\mathbf{p}+\mathbf{q}_1) & \mathbf{0} & \mathbf{0} & \mathbf{0} & \mathbf{0} & \mathbf{0} \\
        T_{2,+}^{\dagger} & \mathbf{0} & H_{G}^{(+)} (\mathbf{p}+\mathbf{q}_2) & \mathbf{0} & \mathbf{0} & \mathbf{0} & \mathbf{0} \\
        T_{3,+}^{\dagger} & \mathbf{0} & \mathbf{0} & H_{G}^{(+)} (\mathbf{p}+\mathbf{q}_3) & \mathbf{0} & \mathbf{0} & \mathbf{0} \\
        T_{1,-}^{\dagger} & \mathbf{0} & \mathbf{0} & \mathbf{0} & H_{G}^{(-)} (\mathbf{p}-\mathbf{q}_1) & \mathbf{0} & \mathbf{0} \\
        T_{2,-}^{\dagger} & \mathbf{0} & \mathbf{0} & \mathbf{0} & \mathbf{0} & H_{G}^{(-)} (\mathbf{p}-\mathbf{q}_2) & \mathbf{0}  \\
        T_{3,-}^{\dagger} & \mathbf{0} & \mathbf{0} & \mathbf{0} & \mathbf{0} & \mathbf{0} & H_{G}^{(-)} (\mathbf{p}-\mathbf{q}_3) \\
    \end{array}
    \right],
    \label{eq24}
\end{equation}
\begin{equation}
    \psi_{\Gamma}^{\dagger}(\mathbf{p})=\left[\mathbf{D}_{\mathbf{p}}^{\dagger}, \mathbf{C}^{\dagger}_{\mathbf{p}+\mathbf{q}_1,+}, 
    \mathbf{C}^{\dagger}_{\mathbf{p}+\mathbf{q}_2,+},
    \mathbf{C}^{\dagger}_{\mathbf{p}+\mathbf{q}_3,+}, \mathbf{C}^{\dagger}_{\mathbf{p}-\mathbf{q}_1,-}, 
    \mathbf{C}^{\dagger}_{\mathbf{p}-\mathbf{q}_2,-},
    \mathbf{C}^{\dagger}_{\mathbf{p}-\mathbf{q}_3,-} \right].
    \label{eq25}
\end{equation}
\end{widetext}

\subsection{Mean field approximation}

NbSe$_2$ exhibits superconductivity regardless of the number of layers, from monolayer to bulk. In contrast, monolayer graphene still be a normal metal down to low temperatures. Thus, we assume that the pairing interaction for superconductivity is present only in the NbSe$_2$ layer. Moreover, simple s-wave superconductivity due to electron-phonon coupling is assumed. Hereafter, we focus on the change of monolayer SC properties by band hybridization, and qualitative properties are expected to be independent of the symmetry of superconductivity.

The paring interaction Hamiltonian for NbSe$_2$ is written as 
\begin{equation}
\mathcal{H}_{\mathrm{int}}=\sum_{\mathbf{k},\mathbf{k}',\sigma_j} V(\mathbf{k},\mathbf{k}') d^{\dagger}_{\mathbf{k},\sigma_1}d^{\dagger}_{-\mathbf{k},\sigma_2}d_{-\mathbf{k}',\sigma_3}d_{\mathbf{k}',\sigma_4}. 
\label{eq26}
\end{equation}
In SC states, the expectation value $\left< d_{-\mathbf{k},\sigma}d_{\mathbf{k},\sigma'}\right>$ is nonzero, and the order parameter is defined by
\begin{equation}
    \Delta_{\sigma_1\sigma_2}(\mathbf{k})=\sum_{\mathbf{k}',\sigma_3,\sigma_4} V(\mathbf{k},\mathbf{k}')\left< d_{-\mathbf{k}',\sigma_3}d_{\mathbf{k}',\sigma_4}\right>.
    \label{eq27}
\end{equation}
In the mean field approximation, the four-body operator of $\mathcal{H}_{\mathrm{int}}$ is approximated by two-body operators, and the pairing interaction Hamiltonian is simplified by the following mean field Hamiltonian:
\begin{equation}
    \mathcal{H}^{(\mathrm{MF})}_{\mathrm{int}}=\sum_{\mathbf{k},\sigma_1,\sigma_2} \Delta_{\sigma_1,\sigma_2}(\mathbf{k}) d^{\dagger}_{\mathbf{k},\sigma_1}d^{\dagger}_{-\mathbf{k},\sigma_2} +h.c..
    \label{eq28}
\end{equation}
Finally, consistent with the assumption of s-wave superconductivity, the pairing interaction has the form,
\begin{equation}
    V(\mathbf{k},\mathbf{k}')=-\dfrac{V_S}{2}(\delta_{\sigma_1,\sigma_4}\delta_{\sigma_2,\sigma_3}-\delta_{\sigma_1,\sigma_3}\delta_{\sigma_2,\sigma_4}).
    \label{eq29}
\end{equation}

The presence of multiple FSs around the $\mathbf{K}_{\xi}$ valleys and the $\Gamma$ point in the NbSe$_2$ layer lead to multigap superconductivity. The simplest formalism deals with independent Cooper pairings on each FS by considering three distinct electron-phonon coupling amplitudes (see Fig.~\ref{fig4}). The order parameter on the $\Gamma$ FS is 
\begin{equation}
    \Delta_{\Gamma}=-V_S^{(\Gamma)} \sum_{\mathbf{p}} \left< d_{\mathbf{p},\uparrow}d_{-\mathbf{p},\downarrow}\right>. 
    \label{eq32} 
\end{equation}
Similarly, the Cooper pairing between the $\mathbf{K}_+$ and $\mathbf{K}_{-}$ valleys are defined by the following order parameters, 
\begin{align}
    \Delta_{\mathrm{outer}}=&-V_S^{(\mathrm{O})}\sum_{\mathbf{p}} \left< d_{\mathbf{p}+\mathbf{K}_{+},\uparrow}d_{-\mathbf{p}+\mathbf{K}_{-},\downarrow}\right>, 
    \label{eq33} \\
    \Delta_{\mathrm{inner}}=&V_S^{(\mathrm{I})}\sum_{\mathbf{p}} \left< d_{\mathbf{p}+\mathbf{K}_{+},\downarrow}d_{-\mathbf{p}+\mathbf{K}_{-},\uparrow}\right>. 
    \label{eq34} 
\end{align}
Note that the two order parameters $\Delta_{\mathrm{inner}}$, $\Delta_{\mathrm{outer}}$
are considered because of spin splitting by the Ising SOC. The schematic image of the above three Cooper pairings is shown in Fig.~\ref{fig4}.

We introduce the following Nambu spinors for each FS: 
\begin{align}
    \Psi_{K,\xi}^{\dagger}(\mathbf{p})=&\left[\psi_{K,\xi}^{\dagger}(\mathbf{p}), \psi_{K,-\xi}^{\mathrm{T}}(-\mathbf{p}) \right],
    \label{eq35}\\
    \Psi_{\Gamma}^{\dagger}(\mathbf{p})=&\left[\psi_{\Gamma}^{\dagger}(\mathbf{p}), \psi_{\Gamma}^{\mathrm{T}}(-\mathbf{p}) \right].
    \label{eq36}
\end{align}
When the FSs of NbSe$_2$ and graphene around the K valleys are hybridized with each other, the Bogoliubov de Gennes (BdG) Hamiltonian describing the SC states is a sum of $\mathcal{H}^{(+)}_{K-K}$, $\mathcal{H}^{(-)}_{K-K}$, and $\mathcal{H}^{(\mathrm{MF})}_{\mathrm{int}}$. If we ignore the FS around $\Gamma$, the pairing interaction Hamiltonian under the mean field approximation becomes
\begin{align}
\mathcal{H}^{(\mathrm{MF})}_{\mathrm{int}}=&\sum_{\mathbf{p}}\Delta_{\mathrm{outer}} \, d^{\dagger}_{\mathbf{p}+\mathbf{K}_{+},\uparrow}d^{\dagger}_{-\mathbf{p}+\mathbf{K}_{-},\downarrow} \notag \\
+&\sum_{\mathbf{p}}\Delta_{\mathrm{inner}} \, d^{\dagger}_{\mathbf{p}+\mathbf{K}_{+},\downarrow}d^{\dagger}_{-\mathbf{p}+\mathbf{K}_{-},\uparrow}+h.c..
\label{eq37}
\end{align}
Therefore, the BdG Hamiltonian is written as
\begin{align}
    \mathcal{H}_{\mathrm{BdG}}^{(K)}=&\sum_{\mathbf{p}}\Psi_{K,+}^{\dagger}(\mathbf{p}) H_{\mathrm{BdG}}^{(+)}(\mathbf{p}) \Psi_{K,+}(\mathbf{p}), \notag \\
    =&\sum_{\mathbf{p}}\Psi_{K,-}^{\dagger}(\mathbf{p}) H_{\mathrm{BdG}}^{(-)}(\mathbf{p}) \Psi_{K,-}(\mathbf{p}),
    \label{eq38}
\end{align}
where
\begin{equation}
    H_{\mathrm{BdG}}^{(\xi)}(\mathbf{p})=
    \begin{bmatrix}
        H_{K-K}^{(\xi)} (\mathbf{p}) & \xi \Delta_K \\
        \xi \Delta_K^{\dagger} & -H_{K-K}^{(-\xi)} (-\mathbf{p})
    \end{bmatrix},
    \label{eq39}
\end{equation}
is a $28 \cross28$ matrix in Nambu space and
\begin{equation}
\Delta_K=
    \begin{bmatrix}
    \begin{matrix}
        0 & \Delta_{\mathrm{outer}}\\
        \Delta_{\mathrm{inner}} & 0\\
    \end{matrix}
         & \mathbf{0} \\
         \mathbf{0} & \mathbf{0} \\
    \end{bmatrix}.
    \label{40}
\end{equation}
It is noteworthy that the sign of the valley $\xi$ used for $\mathcal{H}_{\mathrm{BdG}}^{(K)}$ can be restricted to either $+$ or $-$. We have ignored the FS around the $\Gamma$ point in this formulation, but it can be straightforwardly extended by adding the monolayer BdG Hamiltonian for the FS around $\Gamma$ to $\mathcal{H}_{\mathrm{BdG}}^{(K)}$.

 \begin{figure*}[htbp]
     \centering
     \includegraphics[height=7cm]{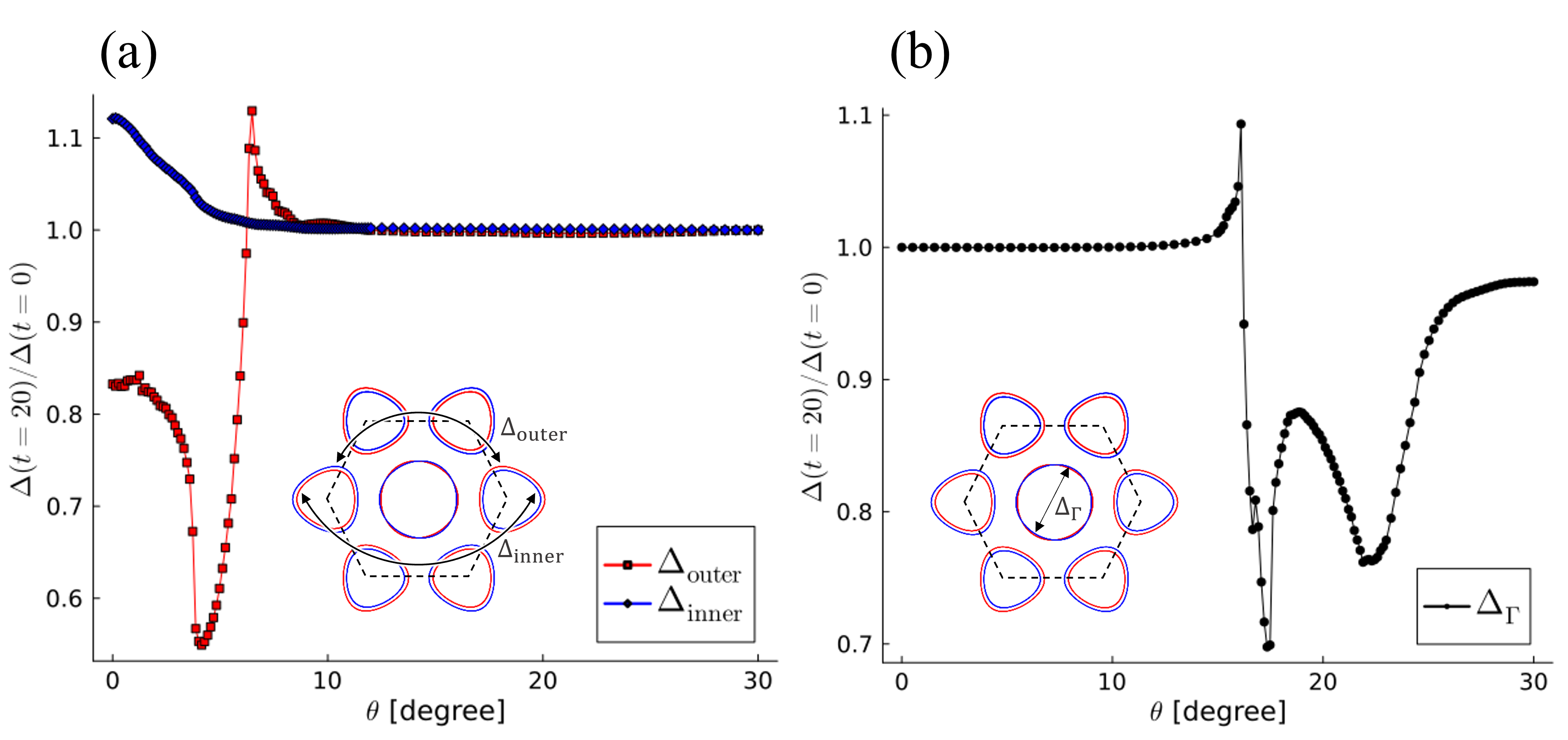}
     \caption{The twist angle dependence of the order parameters (a) $\Delta_{\mathrm{outer}}$, $\Delta_{\mathrm{inner}}$ and (b) $\Delta_{\Gamma}$. The figures show the ratio of the SC gap amplitude to that of the decoupled bilayer equivalent to monolayer NbSe$_2$. The temperature of the system is $T=0$. The SC gap is enhanced in the twisted bilayer when the ratio exceeds $1$, while it is suppressed otherwise.}
     \label{fig5}
 \end{figure*}

In the same way, the BdG Hamiltonian is obtained as follows, when the FSs around the $\Gamma$ point hybridize the FSs of graphene:
\begin{equation}   
\mathcal{H}_{\mathrm{BdG}}^{(\Gamma)}=\frac{1}{2}\sum_{\mathbf{p}}\Psi_{\Gamma}^{\dagger}(\mathbf{p})H_{\mathrm{BdG}}^{(\Gamma)} (\mathbf{p})\Psi_{\Gamma}(\mathbf{p}),
\label{eq41}
\end{equation}
where
\begin{align}
    H_{\mathrm{BdG}}^{(\Gamma)} (\mathbf{p})=& 
    \begin{bmatrix}
        H_{K-\Gamma} (\mathbf{p}) & \Delta_{\Gamma} \Lambda \\
        \Delta_{\Gamma} \Lambda^{\dagger} &- H_{K-\Gamma}^T (-\mathbf{p})
    \end{bmatrix},
 \label{eq42} \\
\Lambda=&
    \begin{bmatrix}
        i \sigma_y & \mathbf{0} \\
        \mathbf{0} & \mathbf{0}
    \end{bmatrix}.
    \label{eq43}
\end{align}

\subsection{Self-consistent gap equation for NbSe$_2$ layer}

In the previous subsection, we constructed the BdG Hamiltonian for the SC states. As the next process, here we shall derive self-consistent gap equations for the SC states~\cite{Sato2017IOP}.

In general, a $2\nu \cross 2\nu$ matrix of the BdG Hamiltonian can be diagonalized as
\begin{align}
    &U^{\dagger}(\mathbf{p})H_{\mathrm{BdG}}(\mathbf{p})U(\mathbf{p}) \notag \\
    &=\mathrm{diag}\left[E_{1}(\mathbf{p}),\cdots E_{\nu}(\mathbf{p}), -E_{1}(\mathbf{-p}),\cdots -E_{\nu}(\mathbf{-p})\right],
    \label{eq44}
\end{align}
with a unitary matrix 
\begin{equation}
    U(\mathbf{p})=
    \begin{bmatrix}
        u_{\alpha}^{(\beta)}(\mathbf{p}) & v_{\alpha}^{(\beta)*}(-\mathbf{p})\\
        v_{\alpha}^{(\beta)}(\mathbf{p}) & u_{\alpha}^{(\beta)*}(-\mathbf{p})
    \end{bmatrix},
    \label{eq45}
\end{equation}
where $\alpha$ and $\beta$ run from $1$ to $\nu$, and the eigenvalue equation
\begin{equation}
    H_{\mathrm{BdG}}(\mathbf{p})
    \begin{bmatrix}
        \mathbf{u}^{(\beta)}(\mathbf{p}) \\ \mathbf{v}^{(\beta)}(\mathbf{p})
    \end{bmatrix}
    =E_{\beta}(\mathbf{p})
    \begin{bmatrix}
        \mathbf{u}^{(\beta)}(\mathbf{p}) \\ \mathbf{v}^{(\beta)}(\mathbf{p})
    \end{bmatrix},
    \label{46}
\end{equation}
is satisfied for the $\beta$-th eigen state. We define new creation/annihilation spinors for Bogoliubov quasiparticles   $\gamma,\gamma^{\dagger}$ as
\begin{equation}
    \begin{bmatrix}
        \psi(\mathbf{p}) \\
        \psi^{\dagger}(-\mathbf{p})^{\mathrm{T}}
    \end{bmatrix}
    =U(\mathbf{p})
    \begin{bmatrix}
        \gamma(\mathbf{p}) \\
        \gamma^{\dagger}(-\mathbf{p})^{\mathrm{T}}
    \end{bmatrix}.
    \label{eq47}
\end{equation}
Here, $\psi$ corresponds to either of $\psi_{K,\xi}$ or $\psi_{\Gamma}$. In this diagonalized basis, the BdG Hamiltonian is written as 
\begin{align}
    \mathcal{H}_{\mathrm{BdG}}=&\dfrac{1}{2}\sum_{\mathbf{p}}\Psi^{\dagger}(\mathbf{p})H_{\mathrm{BdG}}(\mathbf{p})\Psi(\mathbf{p}) \notag \\
    =& \sum_{\nu}\sum_{\mathbf{p}}E_{\nu}(\mathbf{p})\gamma_{\nu}^{\dagger}(\mathbf{p})\gamma_{\nu}(\mathbf{p}).
    \label{eq48}
\end{align}
Thus, the expectation values of $\gamma^{\dagger}\gamma$ are
\begin{equation}
    \left< \gamma_{\nu}^{\dagger}(\mathbf{p})\gamma_{\mu}(\mathbf{p})\right>=f(E_{\nu}(\mathbf{p}))\delta_{\nu,\mu},
    \label{eq49}
\end{equation}
where $f\left(E(\mathbf{p})\right)$ is the Fermi distribution function, and other expectation values such as $\left<\gamma\gamma\right>$ equal 0.

In the remaining part, the self-consistent gap equations are derived. The first and second components of $\psi$ are the operators of electrons on the NbSe$_2$ layer. Thus, from corresponding equations in Eq.~\eqref{eq47}: 
\begin{align}
    \psi_{\mu}(\mathbf{p})&= u_{\mu}^{(\alpha)}(\mathbf{p})\gamma_{\alpha}(\mathbf{p})+v_{\mu}^{(\alpha)*}(-\mathbf{p})\gamma^{\dagger}_{\alpha}(-\mathbf{p}), \label{eq50}\\
    \psi^{\dagger}_{\nu}(-\mathbf{p})&=v_{\nu}^{(\beta)}(\mathbf{p})\gamma_{\beta}(\mathbf{p})+u_{\nu}^{(\beta)*}(-\mathbf{p})\gamma_{\beta}^{\dagger}(-\mathbf{p}),
    \label{eq51}
\end{align}
the gap equations in Eqs.~\eqref{eq32}, \eqref{eq33},  and \eqref{eq34} can be represented by the thermal distribution of Bogoliubov quasiparticles. For instance, in the case of the pairing around the $\Gamma$ point, Eq.~\eqref{eq32} becomes
\begin{equation}
    \Delta_{\Gamma}=-V_{S}^{(\Gamma)}\sum_{\mathbf{p},\alpha}u_{1}^{(\alpha)}(\mathbf{p})v_{2}^{(\alpha)*}(\mathbf{p})\tanh{\frac{E_{\alpha}(\mathbf{p})}{k_{\rm B}T}},
    \label{eq52}
\end{equation}
where we used the expectation values in Eq.~\eqref{eq49}.
Note that $u_{1}^{(\alpha)}(\mathbf{p})$ and $v_{2}^{(\alpha)*}(\mathbf{p})$ are determined by the BdG Hamiltonian and depend on $\Delta_{\Gamma}$. Thus, we can obtain the order parameter $\Delta_{\Gamma}$ by solving the gap equation Eq.~\eqref{eq52} self-consistently. Similarly, the gap equations for the order parameters around the K valleys are obtained by
\begin{align}
    \Delta_{\mathrm{outer}}&=-V_{S}^{(\mathrm{O})}\sum_{\mathbf{p},\alpha}u_{1}^{(\alpha)}(\mathbf{p})v_{2}^{(\alpha)*}(\mathbf{p})\tanh{\frac{E_{\alpha}(\mathbf{p})}{k_{\rm B}T}}, \label{eq53} \\
\Delta_{\mathrm{inner}}&=V_{S}^{(\mathrm{I})}\sum_{\mathbf{p},\alpha}u_{2}^{(\alpha)}(\mathbf{p})v_{1}^{(\alpha)*}(\mathbf{p})\tanh{\frac{E_{\alpha}(\mathbf{p})}{k_{\rm B}T}}. \label{eq54}
\end{align}

In general, only the electronic states near the Fermi level contribute to the SC states. Therefore, the summation of wave numbers is restricted to the low-energy region by introducing an energy cutoff that corresponds to the Debye frequency. In this paper, we set the Debye frequency of NbSe$_2$ as $\hbar \omega_{\mathbf{D}}=30$ meV~\cite{Eremenko2009Physica}, and the sum of wave numbers in Eqs.~\eqref{eq52}, \eqref{eq53}, and \eqref{eq54} are implicitly restricted to the energy window $[-\hbar \omega_{\mathrm{D}},\hbar \omega_{\mathrm{D}}]$ from the Fermi level.

\section{Results} 
\subsection{Twist angle dependence of SC states}
We here compare the superconducting properties in the present twisted bilayer with those of the monolayer NbSe$_2$. 
For a model of the monolayer NbSe$_2$, we set $t=0$~meV where the twisted bilayer is decoupled. We can deduce the properties of the monolayer NbSe$_2$ in this setup and compare the results with those of $t=20$~meV for the twisted bilayer. 

\begin{figure}[htbp]
    \centering
    \includegraphics[width=7cm]{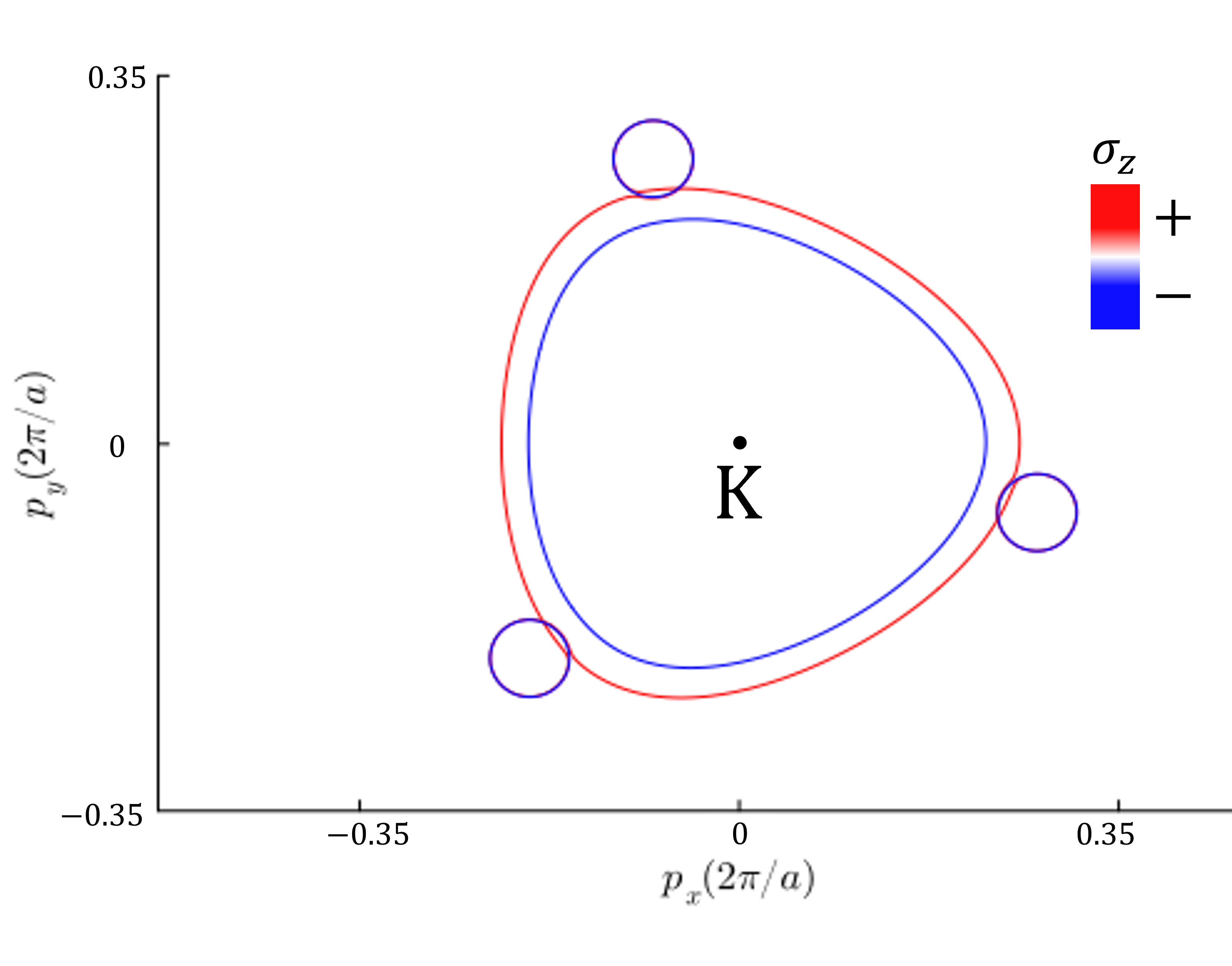}
    \caption{FS of the twisted NbSe$_2$ graphene bilayer around the K valley for $\theta=4^{\circ}$. The outer FS, i.e., the FS of spin up electrons on the NbSe$_2$ layer strongly overlaps with the FSs of graphene. Note that the center of the figure is the K point.}
    \label{fig6}
\end{figure}
\begin{figure}[htbp]
    \centering
    \includegraphics[width=7.5cm]{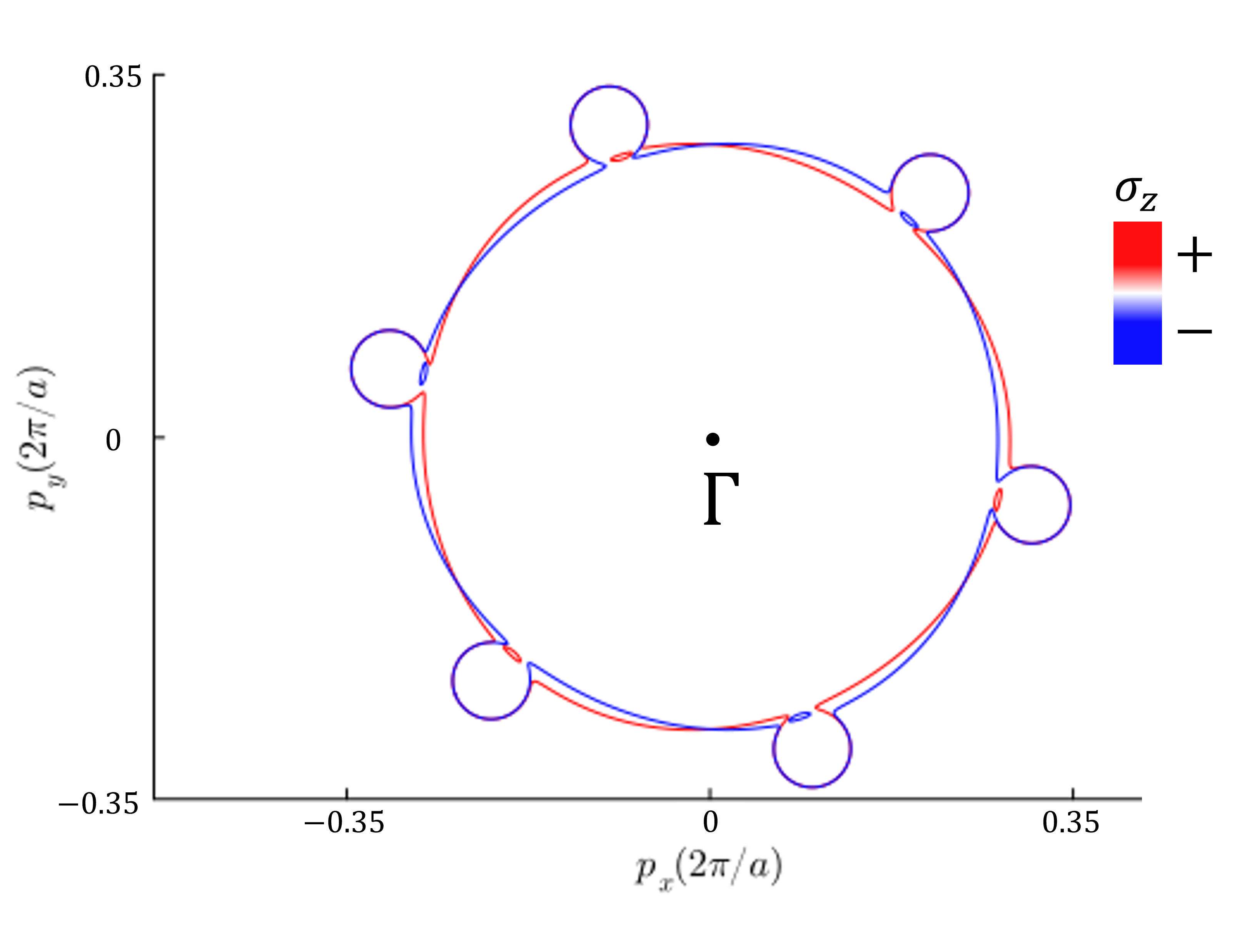}
    \caption{FS of the twisted NbSe$_2$ graphene bilayer around the $\Gamma$ point for $\theta=17^{\circ}$. The figure is depicted with the $\Gamma$ point in the center. 
    }
    \label{fig7}
\end{figure}
\begin{figure}[htbp]
    \centering
    \includegraphics[width=7.5cm]{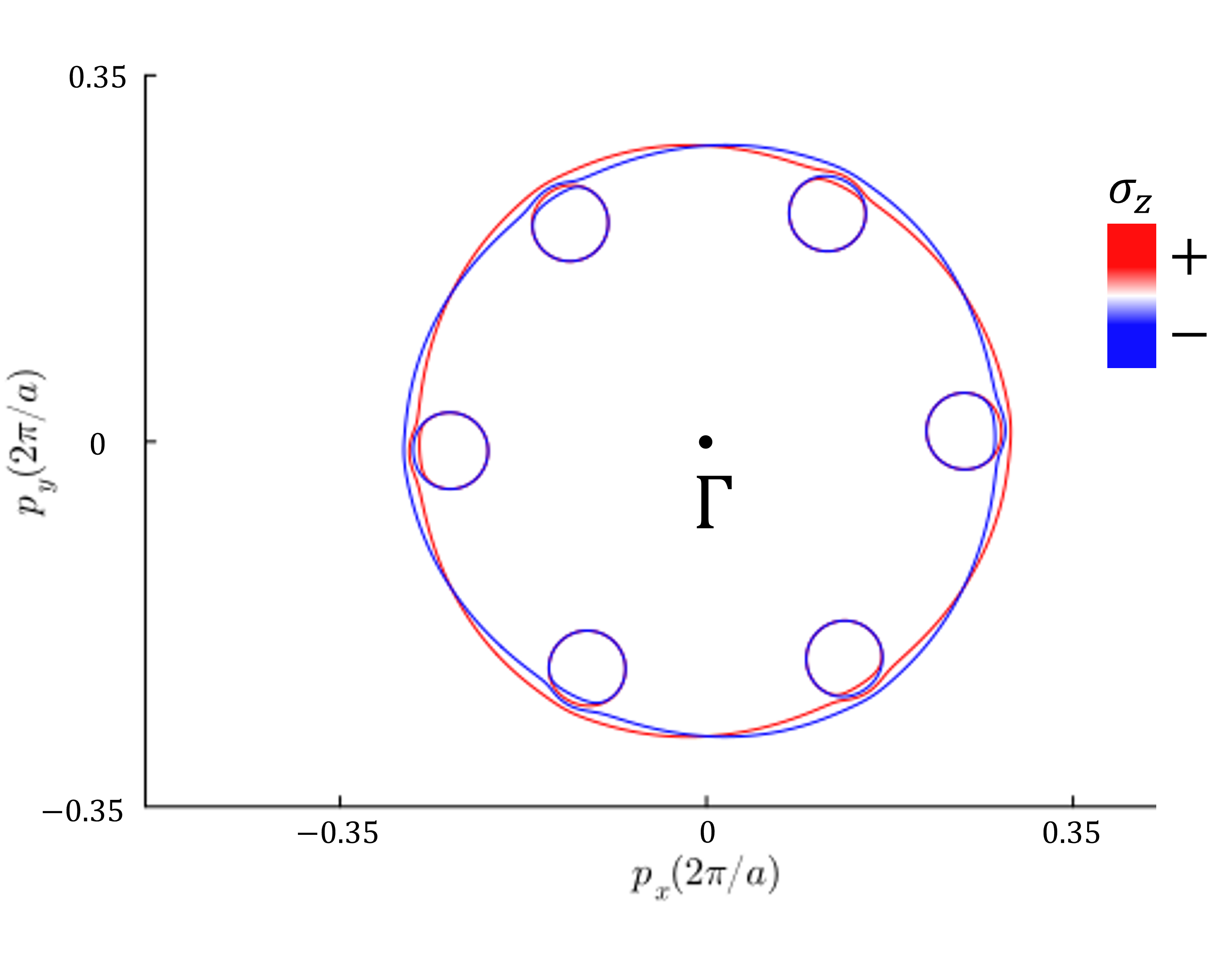}
    \caption{FS of the twisted NbSe$_2$ graphene bilayer around the $\Gamma$ point for $\theta=23^{\circ}$.}
    \label{fig8}
\end{figure}
We first discuss the SC gap functions $\Delta_{\mathrm{outer}}, \Delta_{\mathrm{inner}}$, and $\Delta_{\Gamma}$. Figure~\ref{fig5} shows the twist angle dependence of the gap functions compared to that of the monolayer. The temperature of the system is fixed to zero temperature $T=0$. Here, we set the interaction parameters as $V_{S}^{(\Gamma)}=22$, $V_{S}^{(\mathrm{O})}= 17.5$, $V_{S}^{(\mathrm{I})}=17$ $[\mathrm{eV}\cdot a^2]$ so that the transition temperature in the decoupled limit coincides with the experimental value of monolayer NbSe$_2$, $T_{\rm c}=3$~K. Note that the amplitude of the s-wave gap is plotted only for angles $0^{\circ} \leq \theta \leq 30^{\circ}$ due to the symmetry of the system~\cite{Gani2019PRB}. 
On account of the band hybridization between the SC layer and the normal metallic layer, the SC gap function around the K valleys $\Delta_{\mathrm{outer}}$ diminishes drastically for $\theta \simeq 0^{\circ}$, although it almost coincides with the monolayer for angles around $30^{\circ}$. In contrast, $\Delta_{\Gamma}$ decreases for $\theta \simeq 30^{\circ}$. These behaviors are consistent with the fact that the FSs of graphene are hybridized with the FSs of NbSe$_2$ around the K valleys near $\theta=0^{\circ}$ while with those around the $\Gamma$ point near $\theta=30^{\circ}$. 

\begin{figure*}[htbp]
    \centering
    \includegraphics[height=5.5cm]{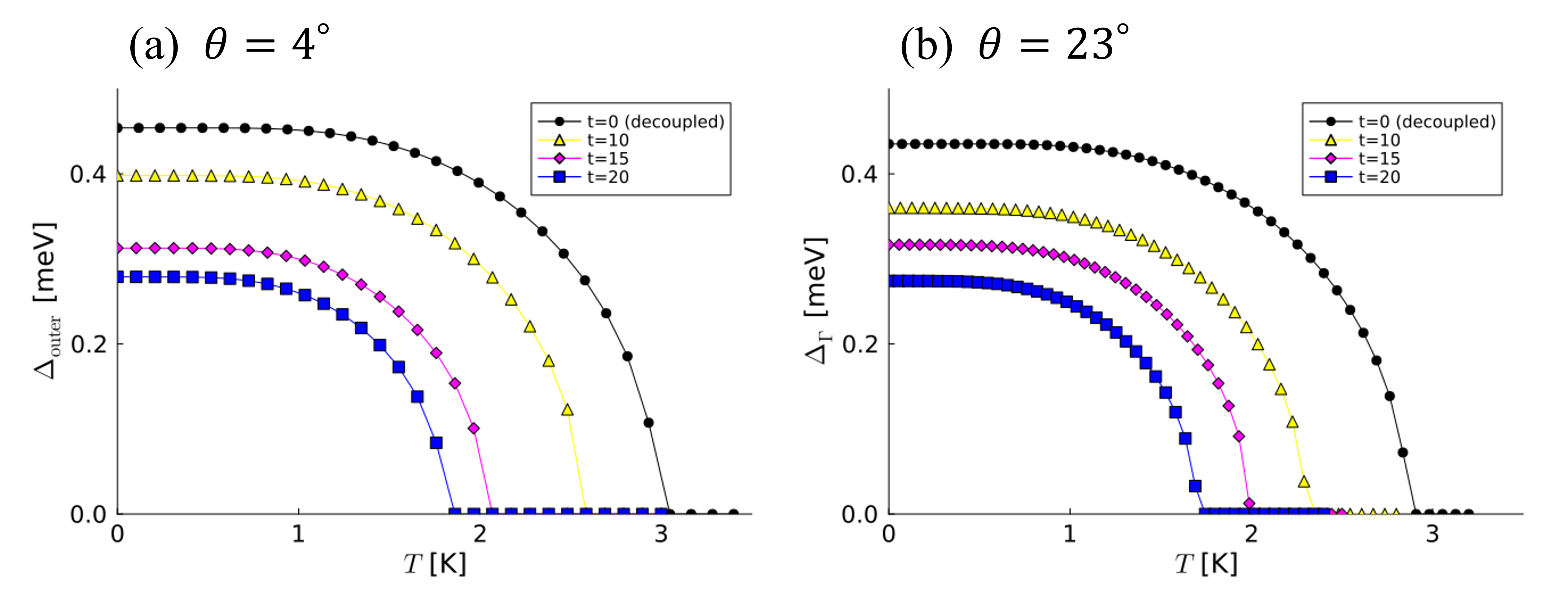}
    \caption{The interlayer coupling dependence of (a) $\Delta_{\mathrm{outer}}$ for $\theta=4^{\circ}$ and (b) $\Delta_{\Gamma}$ for $\theta=23^{\circ}$. The horizontal axis is the temperature $T$. In these twist angles, the SC gap is decreased by the hybridization of NbSe$_2$ and graphene (see Fig.~\ref{fig5}). The SC gaps are suppressed monotonically as the interlayer coupling is enhanced. 
    }
    \label{fig9}
\end{figure*}

The suppression of superconductivity can be attributed to the decrease in the density of states (DOS) near the Fermi level. In the vicinity of the band crossing points of the twisted heterostructure, the graphene states and the NbSe$_2$ states within the finite energy window $t$ hybridize, resulting in a band gap opening. 
The gap opening around the Fermi level decreases the low-energy DOS and suppresses the SC states. Consistent with this view, suppression of the SC gap in NbSe$_2$ is maximized at certain twist angles. That is, $\Delta_{\mathrm{outer}}$ is minimized at $\theta=4^{\circ}$, and $\Delta_{\Gamma}$ is minimized at $\theta=17^{\circ}$ and $23^{\circ}$. For these angles, the FSs of graphene and NbSe$_2$ strongly overlap (see Figs.~\ref{fig6}, \ref{fig7}, and \ref{fig8}), which supports the above interpretation. This result is slightly different from the behavior of the induced SC gap on the graphene layer studied in Ref.~\onlinecite{Gani2019PRB}, although these phenomena have a common origin.

We next focus on the details of SC gap tuning by the bilayer coupling. In the previously shown Fig.~\ref{fig5}, we fixed the temperature $T=0$ K, and the interlayer hopping term was either $t=0$~meV or $20$~meV. Alternatively, we here fix the twist angle as $\theta=4^{\circ}$, $23^{\circ}$, where the SC gap is significantly suppressed, and vary the interlayer coupling $t$. The temperature dependence of the SC gap is shown in Fig.~\ref{fig9}, in which we see that the superconductivity of NbSe$_2$ is monotonically suppressed as the interlayer coupling $t$ increases. This suppression originates from the expansion of the band gap by which the DOS becomes smaller than that of the monolayer system.

Interestingly, from Fig.~\ref{fig5} we can see that the SC gap of NbSe$_2$ can be enhanced by twisting under a certain condition, in contrast to the above cases. For instance, $\Delta_{\mathrm{inner}}$ is larger than that of the decoupled system. Similarly, $\Delta_{\mathrm{outer}}$ and $\Delta_{\Gamma}$ are also enhanced at certain twist angles.
The enhancement of superconductivity by hybridization with a normal metal is counterintuitive. However, we can reveal the mechanism of enhanced superconductivity by discussing the DOS in the normal states (see next subsection).

Finally, we discuss the contrasting behaviors of $\Delta_{\mathrm{inner}}$ and $\Delta_{\mathrm{outer}}$. Although the spin-split K valleys of monolayer NbSe$_2$ have nearly the same SC gaps, they can differ significantly in the twisted bilayer. For example, $\Delta_{\mathrm{outer}}/\Delta_{\mathrm{inner}} < 0.6$ around $\theta=4^{\circ}$. The difference in the SC gap functions can be regarded as parity mixing in Cooper pairs. In the two-band model, the SC gaps of two spin-split FSs are given by $\Delta_{\rm s} \pm \Delta_{\rm t}$, where $\Delta_{\rm s}$ ($\Delta_{\rm t}$) is the order parameter of spin-singlet (spin-triplet) superconductivity~\cite{Bauer2012}. Thus, inequivalence between $\Delta_{\mathrm{inner}}$ and $\Delta_{\mathrm{outer}}$ indicates the significant parity mixing by twisting. We find it interesting that the spin-triplet Cooper pairs can be induced in a controllable way by twisting.

\subsection{Density of states on normal NbSe$_2$ layer}

\begin{figure}[htbp]
    \hspace{-0.5cm}
    \includegraphics[height=5cm]{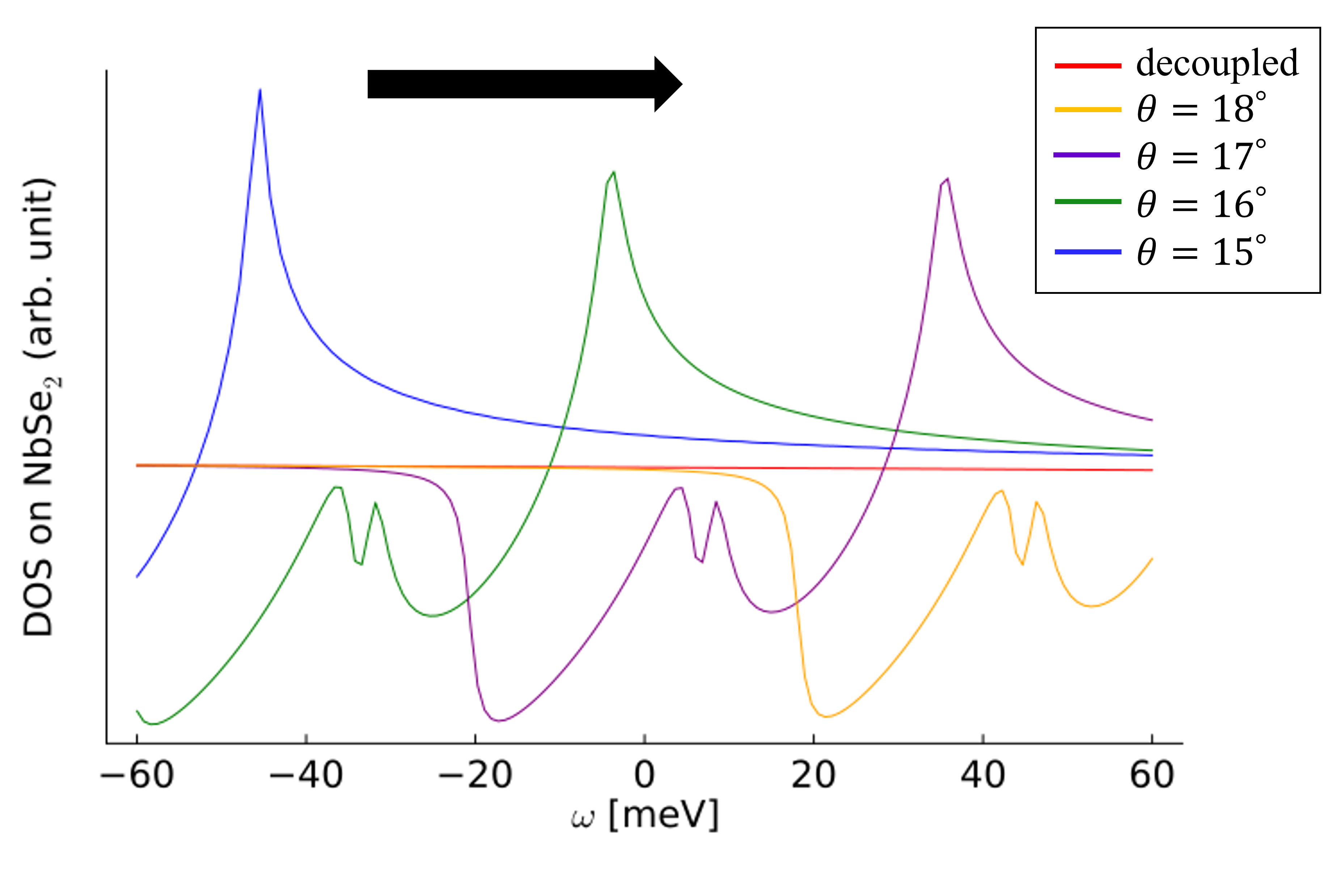}
    \caption{The normal DOS around the $\Gamma$ point calculated by Eq.~\eqref{eq24} for several twist angles. The relationship $D_{\uparrow}(\omega) =D_{\downarrow}(\omega)$ is satisfied, and we here show $D_{\uparrow}(\omega)$ as the normal DOS on the NbSe$_2$ layer. The red line shows the DOS of the decoupled bilayer, i.e., $t=0$~meV, and the others show the results of the twisted bilayers with $t=20$~meV. We see an energy region where the interlayer coupling makes the DOS larger than that of the decoupled layer. The interlayer hybridization suppresses the DOS adjacent to such a region. As the twist angle increases, the DOS shifts to the higher energy side, as illustrated by the black arrow in the figure.}
    \label{fig10}
\end{figure}

\begin{figure*}[ht]
    \centering
    \includegraphics[height=6.0cm]{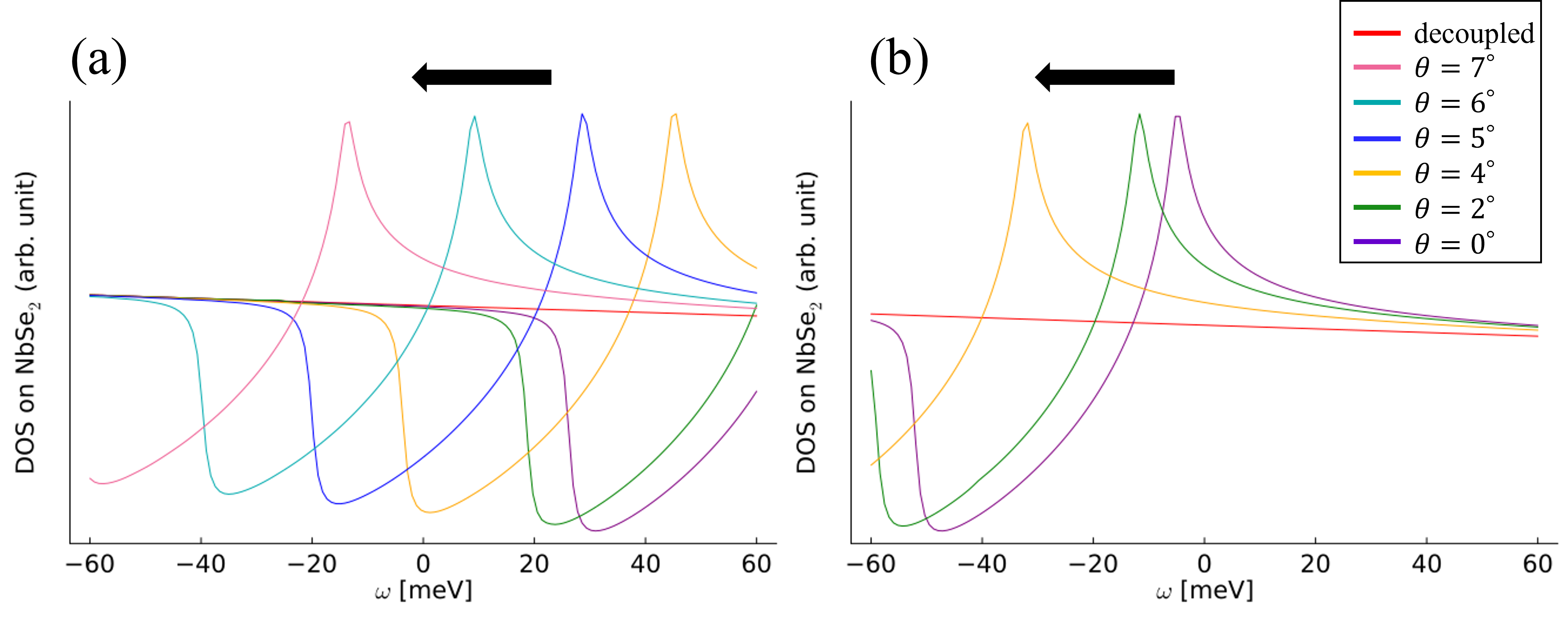}
    \caption{The normal DOS around the K valley calculated by Eq.~\eqref{eq20}. The DOS of (a) spin up electrons $D_{\uparrow}(\omega)$ and (b) spin down electrons $D_{\downarrow}(\omega)$. In contrast to Fig.~\ref{fig10}, the DOS shifts parallel to the lower energy side, as the twist angle increases (represented by the black arrows). 
    }
    \label{fig11}
\end{figure*}

To clarify the enhancement of superconductivity in the twisted bilayer, we calculate the DOS in the normal state. Because NbSe$_2$ is an intrinsic superconductor while graphene is not, we consider the DOS contributed from the NbSe$_2$ layer, not the total DOS of the twisted heterostructure. We then define the spectral function and the DOS contributed from the NbSe$_2$ layer as~\cite{Koshino2015IOP}
\begin{align}
    A_{\sigma}(\mathbf{p},\omega)=&\sum_{\nu} g_{\sigma, \nu}^{(\mathrm{top})}(\mathbf{p})\delta\left(\omega-\epsilon_{\nu}(\mathbf{p})\right), 
    \label{eq55} \\
    D_{\sigma}(\omega)=&\sum_{\mathbf{p}} A_{\sigma}(\mathbf{p},\omega),
    \label{eq56}
\end{align}
where $\sigma=\uparrow, \downarrow$ is an index of spin, $\omega$ is an energy from the Fermi level, $\epsilon_{\nu}(\mathbf{p})$ are energy eigenvalues of the Hamiltonian in Eqs.~\eqref{eq20} and \eqref{eq24}, and $g_{\sigma, \nu}^{(\mathrm{top})}(\mathbf{p})$ is the total amplitudes of spin $\sigma$ on the NbSe$_2$ layer in the eigen state of $\epsilon_{\nu}(\mathbf{p})$. 
Below we show the DOS in the effective Hamiltonian Eqs.~\eqref{eq20} and \eqref{eq24}, in which we focus on the specific FSs around the K point or $\Gamma$ point. Therefore, $D_\sigma(\omega)$ discussed below corresponds to the partial DOS.
For instance, in the case of the effective Hamiltonian for the K valleys Eq.~\eqref{eq20}, the DOS of the spin up/down bands is calculated by substituting the first/second components of eigenvectors for calculating $g_{\sigma, \nu}^{(\mathrm{top})}(\mathbf{p})$. Then,  $D_{\uparrow}(\omega)$ differs from $D_{\downarrow}(\omega)$ by the asymmetry due to spin splitting around the K valley. In contrast, in the Hamiltonian Eq.~\eqref{eq24}, the relationship $D_{\uparrow}(\omega)=D_{\downarrow}(\omega)$ is satisfied because the time reversal symmetry is preserved in the effective Hamiltonian for electrons around the $\Gamma$ point.

We first consider the Hamiltonian Eq.~\eqref{eq24}. This case is simple compared with the case of Eq.~\eqref{eq20}, where the presence of two order parameters $\Delta_{\mathrm{inner}}$ and $\Delta_{\mathrm{outer}}$ makes the relationship between the DOS and SC gaps complicated. Figure~\ref{fig10} shows the DOS $D_{\uparrow}(\omega) \left[=D_{\downarrow}(\omega)\right]$ near the Fermi level calculated from Eqs.~\eqref{eq24} and \eqref{eq56}, in comparison with the decoupled system, i.e., $t=0$~meV.

From Fig.~\ref{fig10}, we notice that the DOS is larger than that of the decoupled system in a certain energy region. This region is located adjacent to the region where the DOS is suppressed by hybridization with graphene as mentioned in the previous subsection. As the twist angle increases, the energy dependence of the DOS shifts almost parallel to the higher energy side due to the shift of momentum that suffers the interlayer band hybridization. This behavior of the normal DOS on the NbSe$_2$ layer is expected to modulate the SC state nonmonotonically. In particular, the substantial enhancement of the DOS near the Fermi level results in a sharp rise of the SC order parameter for $\theta \simeq 16^{\circ}$. In contrast, as the twist angle is subsequently increased, the SC order parameter is decreased by a steep decrease of the DOS at the Fermi level (Fig.~\ref{fig10}). This predicted behavior agrees well with the result of Fig.~\ref{fig5}. These results suggest that we can control suppression and enhancement of the SC state by modifying the normal DOS through graphene substrates.

We next consider the Hamiltonian for the K valleys Eq.~\eqref{eq20}. The gap function $\Delta_{\mathrm{outer}}$ is related to the spin up band, while $\Delta_{\mathrm{inner}}$ is related to the spin down band in our definition. Since $D_{\uparrow}(\omega) \ne D_{\downarrow}(\omega)$, we show $D_{\uparrow}(\omega)$ and $D_{\downarrow}(\omega)$ in Figs.~\ref{fig11}(a) and  \ref{fig11}(b), respectively. Similarly to the case of Eq.~\eqref{eq24} and Fig.~\ref{fig10}, we see the energy regions where the DOS is suppressed or enhanced by band hybridization.
On the other hand, in contrast to Fig.~\ref{fig10}, the DOS shifts to the lower energy side as the twist angle increases. This dissimilarity stems from the contrasting motion of graphene FSs in the Brillouin zone of NbSe$_2$ with tuning the twist angle. Near $\theta \simeq 0^{\circ}$, the graphene FSs move away from the NbSe$_2$ FSs around the K valley as $\theta$ increases. In contrast, increasing the twist angle around $\theta \simeq 16^{\circ}$ makes the graphene FSs closer to the NbSe$_2$ FSs around the $\Gamma$ point. These behaviors result in the qualitatively different twist angle dependence of the DOS.
Analysis of the DOS is consistent with the twist angle dependence of the SC gap functions in Fig.~\ref{fig5}.
In Fig.\ref{fig11}(a), the DOS at the Fermi level is most significantly suppressed near $\theta \simeq 4^{\circ}$, where the SC gap function $\Delta_{\mathrm{outer}}$ shows the minimum. 
In contrast, Fig.~\ref{fig11}(b) reveals that the region with decreased $D_{\downarrow}(\omega)$ can never encompass the Fermi level for arbitrary twist angles. This indicates that the gap function $\Delta_{\mathrm{inner}}$ of the inner FS never becomes smaller than the decoupled case, and this is indeed seen in Fig.~\ref{fig5}. 
The contrasting behavior of the gap functions $\Delta_{\mathrm{inner}}$ and $\Delta_{\mathrm{outer}}$ comes from the non-equivalence between the up and down spins originating from the Ising SOC. The non-equivalence is usually small because the SOC is smaller than the Fermi energy. However, in the twisted bilayer, the graphene's electrons selectively hybridize with the up spin electrons around the K valley of NbSe$_2$, which amplifies the difference of $D_{\uparrow}(\omega)$ and $D_{\downarrow}(\omega)$ and accordingly the non-equivalence of $\Delta_{\mathrm{inner}}$ and $ \Delta_{\mathrm{outer}}$. In other words, the effects of the Ising SOC, such as the parity mixing in Cooper pairs, can be enhanced by the twisted heterostructures.

\begin{figure}[htbp]
    \centering
    \includegraphics[width=9cm]{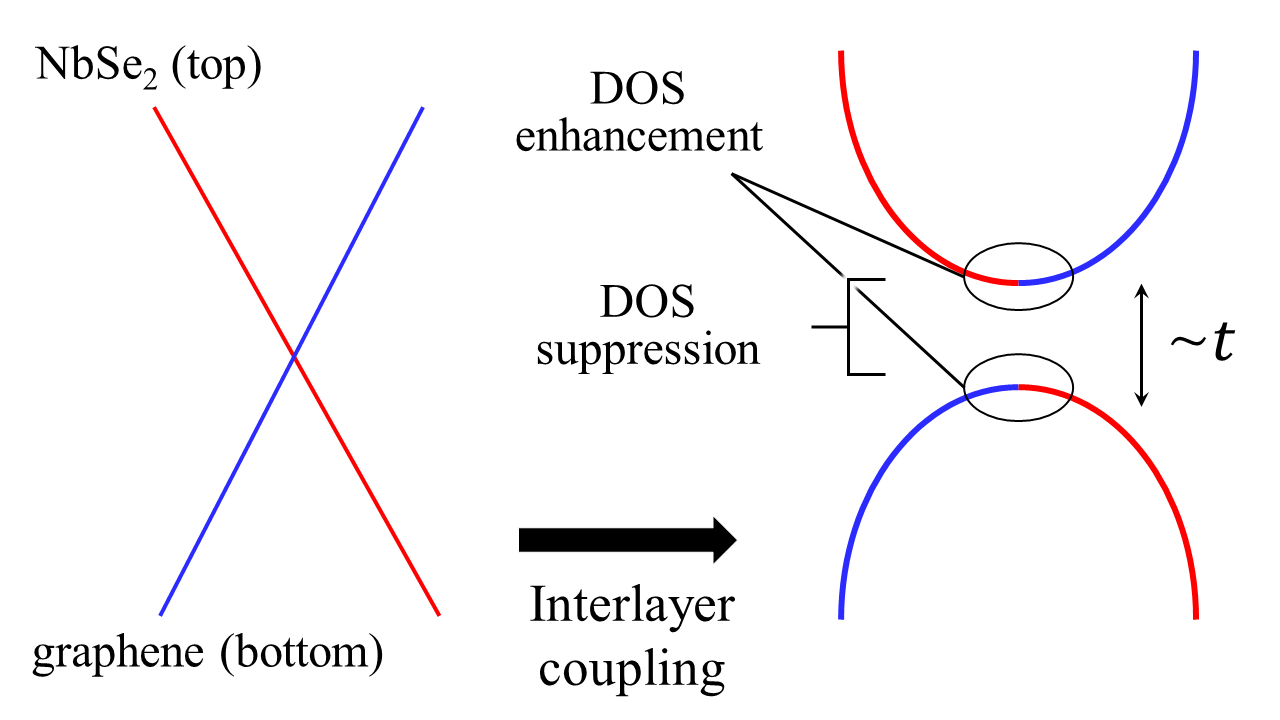}
    \caption{A schematic image of the DOS modulation by band hybridization. The red and blue lines symbolize the energy bands of the NbSe$_2$ and graphene layers, respectively. Due to the interlayer coupling, a band gap of magnitude $t$ opens. The band gap suppresses the DOS in the in-gap region. In conjunction with this band gap opening, the DOS at the edge of the energy band is increased (black circles in the figure).}
    \label{fig12}
\end{figure}

\begin{figure*}[t]
    \centering
    \includegraphics[width=17cm]{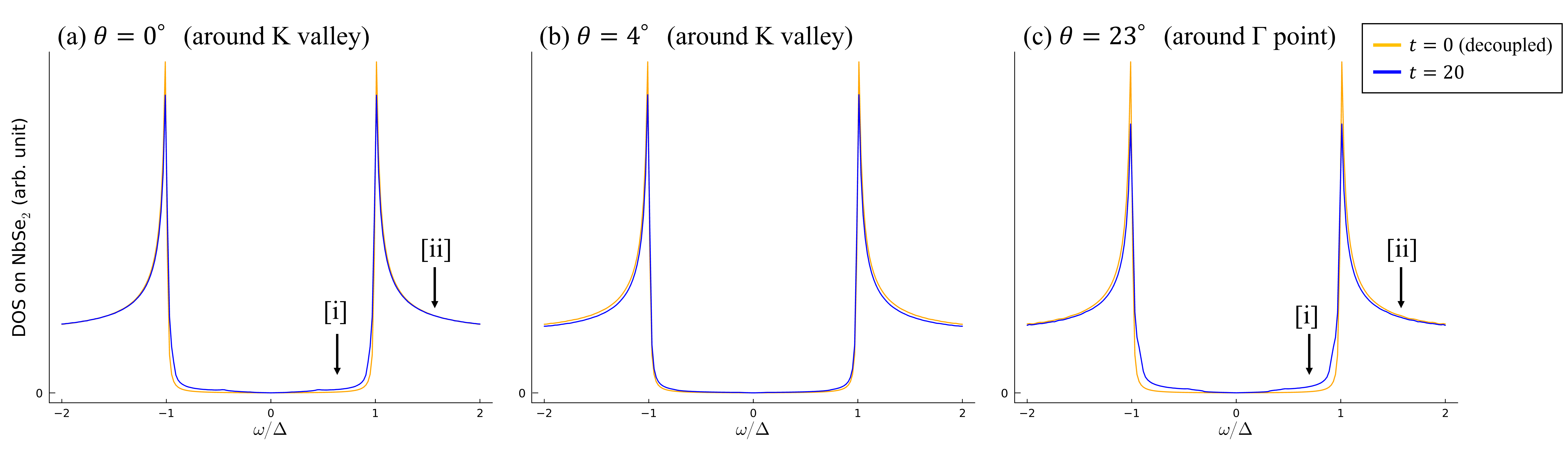}
    \caption{Calculated DOS on the superconducting NbSe$_2$ layer for (a) $\theta=0^{\circ}$, (b) $\theta=4^{\circ}$, and (c) $\theta=23^{\circ}$. We ignore the FSs of NbSe$_2$ around the $\Gamma$ point in (a) and (b) because they are not affected by the interlayer coupling. Similarly, the FSs around the K valleys are ignored in (c). In all the cases, the DOS is modulated compared to that of the monolayer, i.e., the decoupled system. In particular, the DOS becomes larger inside the coherence peaks in common. 
    We also depict two energy values [i] $\omega=0.7\Delta$ and [ii] $\omega=1.5\Delta$ for later Figs.~\ref{fig14} and \ref{fig15}.}
    \label{fig13}
\end{figure*}

\begin{figure*}[ht]
    \centering
    \includegraphics[width=12.7cm]{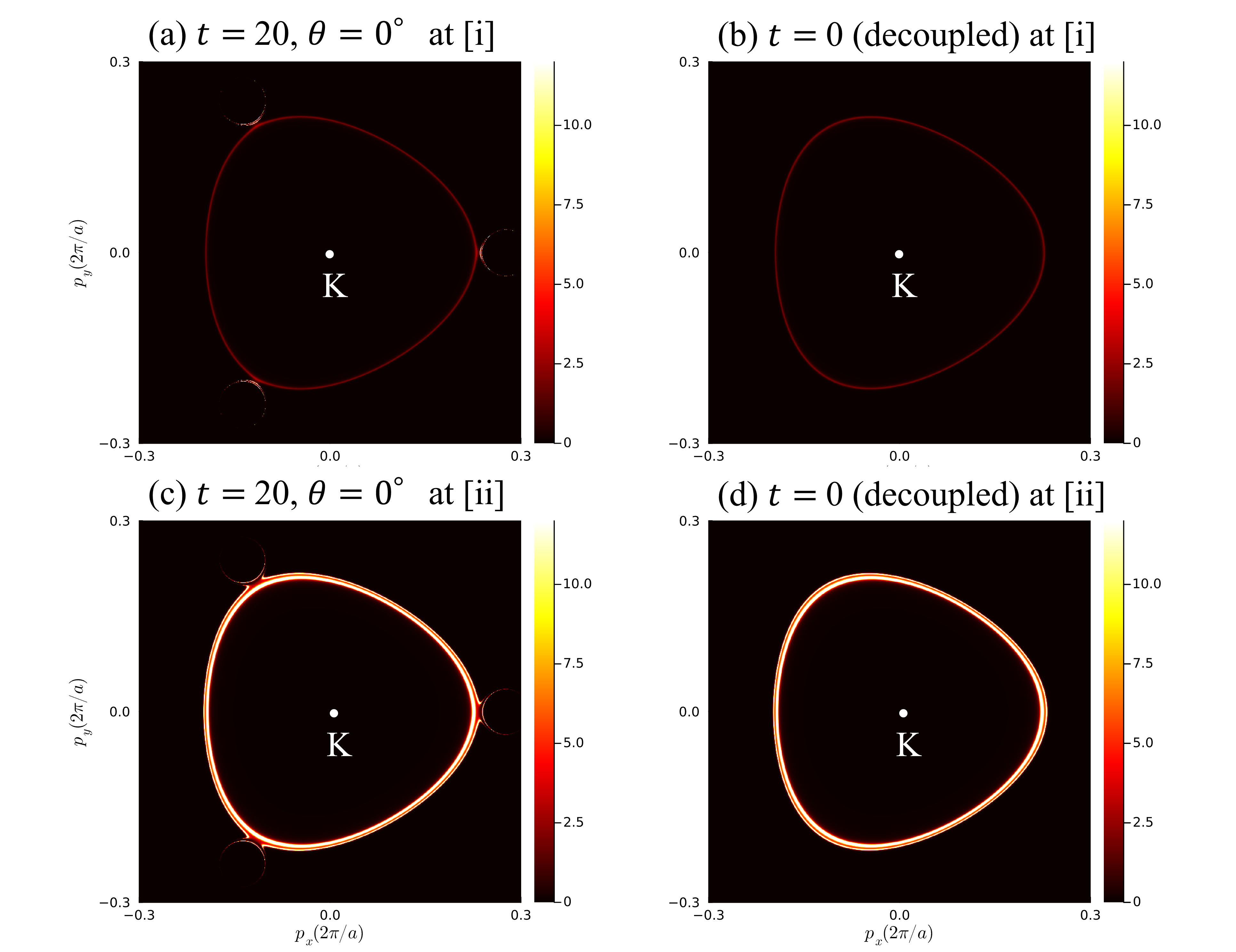}
    \caption{Spectral function around the K valley for $\theta=0^{\circ}$. The upper (lower) figures show the spectrum inside (outside) the SC coherence peaks. The left (right) figures show the spectrum of the twisted bilayer (decoupled layer). Whereas the bright lines originating from the NbSe$_2$ energy band are clearly visible in (c) and (d), they are faintly visible in (a) and (b). However, in (a), the hot spots of low-energy states exist near the crossing points of the NbSe$_2$ and graphene energy bands.}
    \label{fig14}
\end{figure*}

\begin{figure*}[t]
    \centering
    \includegraphics[width=18.5cm]{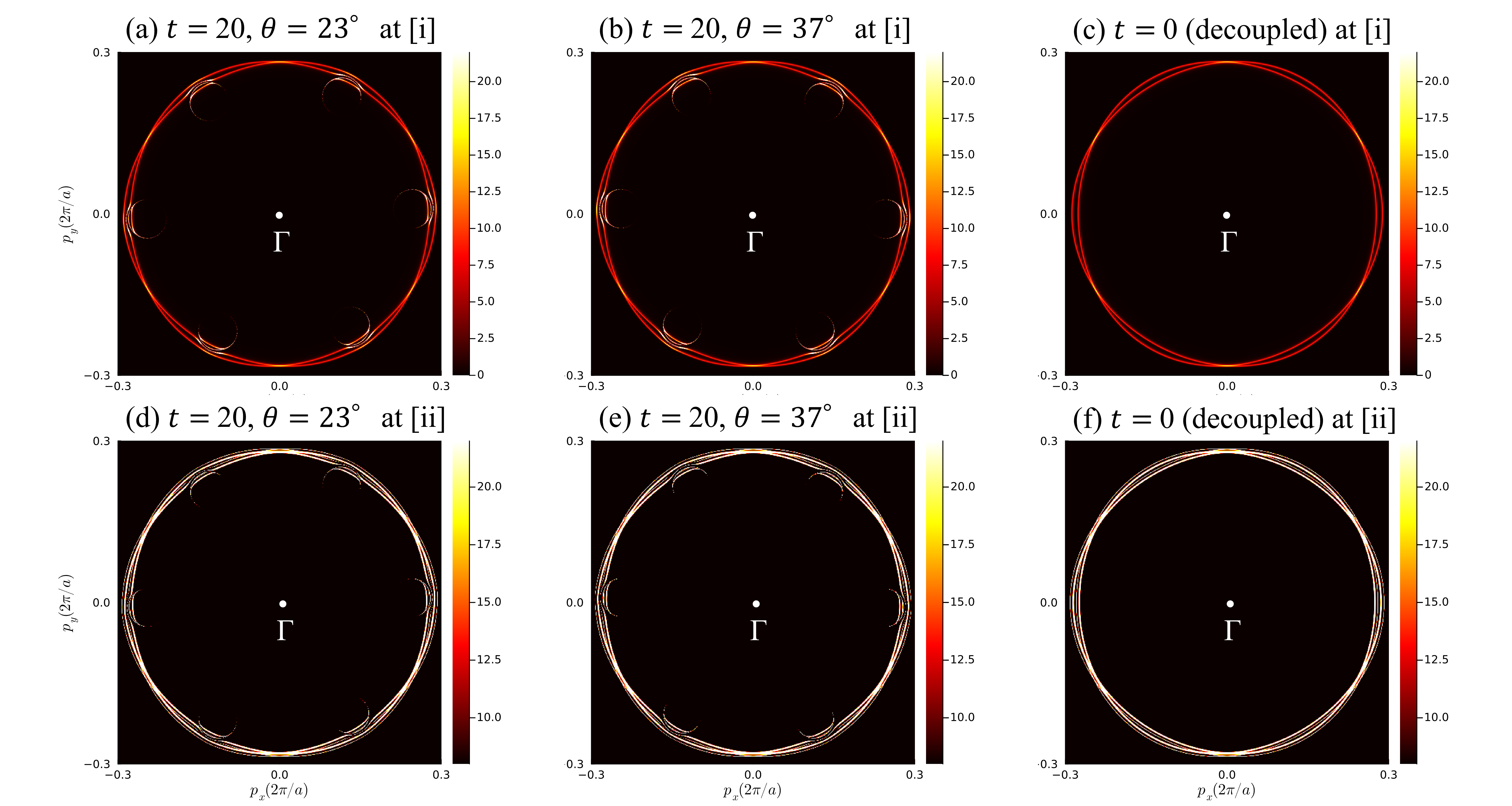}
    \caption{Spectral function around the $\Gamma$ point. The upper (lower) figures are the spectrum inside (outside) the coherence peaks. (a), (b), (d), and (e) show the results of the twisted NbSe$_2$ graphene bilayer, while (c) and (f) are the results of the decoupled case with $t=0$~meV. In (a) and (d) the twist angle is set $\theta=23^{\circ}$ corresponding to Fig.~\ref{fig13}(c). We also show the spectral function for $\theta=37^{\circ}$ in (b) and (e), which are found to be the mirror reflection of (a) and (d).}
    \label{fig15}
\end{figure*}

To conclude this section, we consider the cause of the characteristic shape of the DOS. In all the cases, the normal DOS shows the common feature that the energy region with enhanced DOS is adjacent to one with suppressed DOS. We show a schematic image in Fig.~\ref{fig12} depicting the relationship between the band hybridization and the normal DOS. The red and blue lines symbolically represent the energy band of the NbSe$_2$ and graphene layers. In the presence of the interlayer coupling, a minigap opens at the band crossing point. This leads to the formation of an empty region, which corresponds to the energy region with suppressed DOS in Figs.~\ref{fig10} and \ref{fig11}. In conjunction with this minigap formation, the saddle points at the edge of the energy bands create van Hove singularities~\cite{Moon2013PRB,Brihuega2012PRL,Havener2014Nanolett,Jones2020AdvMater}. These van Hove singularities enhance the normal DOS, leading to the peaks in the DOS that we see in Figs.~\ref{fig10} and \ref{fig11}. The van Hove singularities are inevitably adjacent to the band gap, and therefore, the energy region with the suppressed normal DOS and that with the enhanced DOS appear adjacent. When we vary the twist angle, the position of the band-crossing point shifts to either a higher or lower energy side. 
This shift is the cause of the parallel shift of the normal DOS.

\subsection{Low-energy Bogoliubov quasiparticles on NbSe$_2$}
To obtain further insights into the SC states in twisted heterostructures, we here investigate the properties of Bogoliubov quasiparticles. 
We have assumed simple momentum-independent s-wave gap functions. However, the SC gap of Bogoliubov quasiparticles is not uniform in the momentum space. To study the SC gap in detail, we calculate the spectral function and the DOS in the SC state. 
In this section, we intentionally set $\Delta_{\Gamma},\Delta_{\mathrm{inner}}, \Delta_{\mathrm{outer}}=10$~meV and diagonalize the BdG Hamiltonian Eqs.~\eqref{eq38} and \eqref{eq41}. 

Similarly to the normal state, 
the spectral function and the DOS contributed from the  superconducting NbSe$_2$ layer are given by
\begin{align}
    A^{(\mathrm{SC})}(\mathbf{p},\omega)=&\sum_{\nu} \Big[\left(|u_{1}^{(\nu)}(\mathbf{p})|^2+|u_{2}^{(\nu)}(\mathbf{p})|^2\right)\delta\left(\omega-E_{\nu}(\mathbf{p})\right)  \notag \\
    +&\left(|v_{1}^{(\nu)}(\mathbf{p})|^2+|v_{2}^{(\nu)}(\mathbf{p})|^2\right)\delta\left(\omega+
    E_{\nu}(\mathbf{p})\right) \Big],
    \label{eq57} \\
    D^{(\mathrm{SC})}(\omega)=&\sum_{\mathbf{p}} A^{(\mathrm{SC})}(\mathbf{p},\omega),
    \label{eq58}
\end{align}
where we represent the total amplitudes of the NbSe$_2$ layer by $u_{\alpha}^{(\nu)}(\mathbf{p})$ and $v_{\alpha}^{(\nu)}(\mathbf{p})$ ($\alpha=1,2$) in Eq.~\eqref{eq45}, and we now consider the physical quantities by summing for spin $\sigma=\uparrow,\downarrow$.
Note that $D^{(\mathrm{SC})}(\omega)$ is the partial DOS on the NbSe$_2$ layer and not the total DOS including the graphene states, which were studied in Ref.~\onlinecite{Gani2019PRB}.
The scanning tunneling spectroscopy from the top NbSe$_2$ layer measures the partial DOS rather than the total DOS.

Figure~\ref{fig13} shows the results of $D^{(\mathrm{SC})}(\omega)$ calculated by $\mathcal{H}_{\mathrm{BdG}}^{(K)}$ for $\theta=0^{\circ}, 4^{\circ}$ and by $\mathcal{H}_{\mathrm{BdG}}^{(\Gamma)}$ for $\theta=23^{\circ}$ in comparison with the decoupled system. We notice that the DOS in the SC states is modified by the graphene substrate. 
In particular, the DOS inside the coherence peaks is larger than the decoupled system due to the interlayer band hybridization, which is especially noticeable in Figs.~\ref{fig13}(a) and \ref{fig13}(c). This result exemplifies that the low-energy Bogoliubov quasiparticles emerge on the NbSe$_2$ layer by the interlayer coupling.

To clarify the characteristics of the low-energy quasiparticles, we show the calculated spectral function $A^{(\mathrm{SC})}(\mathbf{p},\omega)$ around the K valley and the $\Gamma$ point in Figs.~\ref{fig14} and \ref{fig15}. In these figures, we set [i] $\omega=0.7\Delta$ or [ii] $\omega=1.5\Delta$ where $\Delta=\Delta_{\Gamma}, \Delta_{\mathrm{inner}}$ or $\Delta_{\mathrm{outer}}$. We also show the spectral function in the decoupled system. A common feature is that the bright white lines originating from the NbSe$_2$ energy band are visible outside the coherence peaks in Figs.~\ref{fig14}(c),(d) and \ref{fig15}(d)-(f). On the other hand, they are only faintly visible inside the coherence peaks due to the broadening, as represented by the red lines in Figs.~\ref{fig14}(a),(b) and \ref{fig15}(a)-(c). However, in the twisted NbSe$_2$ graphene heterostructures, the high-intensity spots appear near the crossing points of the NbSe$_2$ and graphene energy bands. 
This means that the low-energy Bogoliubov quasiparticles emerge on the NbSe$_2$ layer through band hybridization. Indeed, these low-energy states disappear in the absence of the interlayer coupling, i.e., in the monolayer NbSe$_2$ [see Figs.~\ref{fig14}(b) and \ref{fig15}(c)].

It is worth mentioning that except for $\theta=0^{\circ}$ and $30^{\circ}$, the spectral functions $A^{(\mathrm{SC})}(\mathbf{p},\omega)$ break the in-plane mirror symmetry, which reflect the mirror symmetry breaking in real space. In other words, the twisted bilayer with twist angle $\theta$ and that with $-\theta$ are mirror images of each other~\cite{Zhu2024NatMater}. Moreover, the twisted bilayer with $\theta$ is equivalent to the one with $60^{\circ}-\theta$ due to the symmetry in reciprocal space. In fact, the spectral function for $\theta=37^{\circ}$ shown in Figs.~\ref{fig15}(b) and \ref{fig15}(e) are the mirror reflection of Figs.~\ref{fig15}(a) and \ref{fig15}(d), respectively. The in-plane mirror symmetry breaking by twisting combined with the out-of-plane mirror symmetry breaking by the heterostructure realizes a chiral crystalline structure in the real space. Accordingly, our results have verified the Bogoliubov quasiparticles with a chiral structure in momentum space.

\section{Conclusion}
In this paper, we have shown that a combination of a $\mathrm{moir\acute{e}}$ structure and metallic substrate effects tunes the monolayer superconductivity drastically. In particular, the twisted NbSe$_2$ graphene heterostructure has been considered, assuming the situation of a monolayer NbSe$_2$ stacked with a twist on a doped graphene substrate.
By considering the generalized Umklapp process, we have developed a model applicable to a wider range of systems than the previous study. Our analysis of the model elucidated the SC states in detail.

In the numerical calculation, we revealed that the SC states on the NbSe$_2$ layer, i.e., the order parameters are modified by varying the twist angle. The order parameters are strongly suppressed when the FSs of NbSe$_2$ and graphene layers maximally overlap. This result indicates that the modulation of the SC states originates from the interlayer band hybridization between a monolayer superconductor and a monolayer normal metal. Furthermore, we uncovered that superconductivity can also be enhanced by slightly changing the twist angle. The interlayer band hybridization creates a region where the DOS on the NbSe$_2$ layer is increased. As a result, the SC gap and the transition temperature increase when the enlarged DOS is located at the Fermi level.
We also showed that the interlayer hybridization selectively enhances or suppresses the SC gap on the spin-split K valleys of NbSe$_2$ due to the Ising SOC and effectively induces significant parity mixing in Cooper pairs.

We next studied the SC quasiparticles in detail.  
The low-energy Bogoliubov quasiparticles originating from the graphene layer have been found to emerge on the NbSe$_2$ layer. That is, through the interlayer coupling the electronic structures seep out of the bottom graphene layer on the top NbSe$_2$ layer. These low-energy quasiparticles reflect the in-plane mirror symmetry breaking, consistent with the recent observation of chiral Bogoliubov quasiparticles in the twisted NbSe$_2$ graphene heterostructure~\cite{Naritsuka_private}.

Our results support the usefulness of designing monolayer superconductivity by twist and substrate effects. This perspective unveils a way to realize unconventional superconductivity and explore the application of functional superconductors. Moreover, our model can be extended to study other TMDs, other material groups, and even cases where the crystal structure of each layer is different. It is expected to be fruitful in exploring a wide range of systems in future work.

\section*{acknowledgement}
We appreciate Masahiro Naritsuka, Tadashi Machida, and Tetsuo Hanaguri for fruitful discussions from an experimental viewpoint. We also thank fruitful discussions with Michiya Chazono, Ryotaro Sano, Shin Kaneshiro, Hiroto Tanaka, and Koki Shinada. This work was supported by JSPS KAKENHI (Grant Nos. JP21K18145, JP22H01181, JP22H04933, JP23K22452, JP23K17353, JP24H00007).

\bibliography{ref}
\end{document}